\documentclass[12pt,a4paper]{article}

\usepackage{jheppub}
\usepackage{graphicx}
\usepackage{amssymb}
\usepackage{amsmath}
\usepackage{bm}
\usepackage[dvipsnames]{xcolor}
\usepackage{subfigure}
\usepackage{amsfonts}
\usepackage{bm}
\usepackage{mathtools}
\usepackage{nicefrac}
\usepackage{tikz-cd}
\usepackage{siunitx}
\usepackage[makeroom]{cancel}
\usepackage[normalem]{ulem}
\usepackage{appendix}

\definecolor{Orange}{cmyk}{0,0.61,0.87,0}
\definecolor{JungleGreen}{cmyk}{0.99,0,0.52,0}
\definecolor{OliveGreen}{cmyk}{0.64,0,0.95,0.40}
\definecolor{Brown}{cmyk}{0,0.81,1,0.60}
\definecolor{RoyalBlue}{cmyk}{0.71,0.53,0,0.12}

\usepackage{amsthm}

\theoremstyle{definition}

\newcommand{\amp}{\mathcal{M}}
\newcommand{\Mp}{M_{\rm Pl}}
\newcommand{\Mm}{\bar{\Lambda}}
\newcommand{\gm}{\bar{g}}

\usepackage{fullwidth}
\usepackage{caption}
\usepackage{float}

\allowdisplaybreaks[1]

\makeatletter
\renewcommand\@dotsep{200}   
\makeatother

\usepackage{soul}
\usepackage[noabbrev]{cleveref}

\numberwithin{equation}{section}

\title{Discontinuity at the Pole}

\author[a,b]{Dominik Haslehner}
\author[b]{Elena Kaiser}
\author[c]{Javi Serra}
\author[b]{Andreas Weiler}

\affiliation[a]{Max Planck Institute for Physics,
Boltzmannstr.~8, 85748 Garching, Germany}

\affiliation[b]{Technical University of Munich, Physics Department,
James-Franck-Str.~1, 85748 Garching, Germany}

\affiliation[c]{Instituto de F\'{\i}sica Te\'orica UAM/CSIC,
Madrid 28049, Spain}

\emailAdd{dominik.haslehner@tum.de}
\emailAdd{elena.kaiser@tum.de}
\emailAdd{javi.serra@ift.csic.es}
\emailAdd{andreas.weiler@tum.de}

\abstract{
The exchange of a particle whose mass is taken to zero
need not reproduce the exchange of a particle that is strictly massless from the outset. We derive a general on-shell diagnostic of the well-known discontinuity between the exchange of a massless particle and the massless limit of a massive-particle exchange, valid for a mediator of arbitrary integer spin $S$. Our main observation is that this discontinuity can be identified directly at the level of the pole residue.  We find that, in four dimensions, the residue is governed
by a Chebyshev polynomial in the massless case and by a Legendre polynomial in the massless limit of the massive case. For $S = 2$, this reproduces the van Dam-Veltman-Zakharov discontinuity. For general $S$, it provides a simple and compact higher-spin analogue from on-shell factorisation. 
The discontinuity is
probed, in principle, by the leading post-Newtonian correction to the non-relativistic potential between two massive scalar bodies.
}

\begin{document}
\maketitle
\flushbottom


\numberwithin{equation}{section}

\section{Introduction}\label{sec:introduction}

Massive particles with spin are intrinsically different from massless ones: the number of physical degrees of freedom is different. It therefore comes as no surprise that this affects the force mediated by a spinning particle, even when its mass is taken to be parametrically small compared to the energy of the process. The classic example is the van Dam--Veltman--Zakharov discontinuity: in flat space, the $m\to0$ limit of massive spin-two exchange retains a scalar polarisation whose effect does not disappear from physical amplitudes \cite{vanDam:1970vg,Zakharov:1970cc,Boulware:1972yco,Vainshtein:1972sx,Hinterbichler:2011tt}. This raises a broader question. Given a massive spin-$S$ particle exchanged between scalar sources, what part of the scattering amplitude knows whether the exchanged state was strictly massless or obtained as the massless limit of a massive representation?

In this paper, we answer this question at the level of on-shell factorisation. We consider scalar scattering mediated by a single integer-spin particle in four dimensions and compare two objects: the residue of a strictly massless spin-$S$ exchange pole, and the $m\to 0$ limit of the residue of a massive spin-$S$ exchange pole. The residue is the natural quantity to compare because it is fixed by on-shell three-point amplitudes and little-group covariance. It is insensitive to gauge choices, field redefinitions, and contact terms.

We therefore argue that the relevant comparison is naturally formulated at the level of the \emph{factorisation residue} in the exchange channel. On the strictly massless side, gluing the corresponding three-point amplitudes sums only over the helicities $h=\pm S$ and leads to a Chebyshev structure $T_S$. By contrast, taking the $m\to 0$ limit of the massive spin-$S$ exchange retains additional helicity components with fixed relative weights, leading instead to a
Legendre structure $P_S$. The mismatch between the two residues furnishes a clean on-shell diagnostic of the discontinuity between a strictly massless exchange and the massless limit of a massive one. Thus, the discontinuity is captured by the universal replacement
\[
T_S(y)\quad \longleftrightarrow\quad P_S(y),
\]
where \(y\) is the invariant angular variable on the exchange pole. 

The residue-level diagnostic has direct physical implications. After matching the leading static interaction between two non-relativistic scalar sources, the first robust long-range difference between the Chebyshev and Legendre structures appears at relative order \(v^2\). This correction is governed by the difference of the near-forward slopes,
\[
T'_S(1)-P'_S(1) = \frac{S(S-1)}{2}.
\]
This coefficient vanishes for \(S=0,1\) and is non-zero for \(S\ge2\), matching the expectation that the first non-trivial discontinuity occurs at spin two.

The comparison made here is restricted to the factorisation residue, where the on-shell data are well defined. Extending the comparison away from the pole would require specifying an off-pole numerator and possible contact terms, which are in general singular as $m \to 0$ for $S>2$ \cite{Berends:1979je}. The on-shell formalism allows us to neatly disentangle these two different aspects of higher-spin amplitudes.
Furthermore, we are aware that, for $S > 2$, this comparison is not fully physical, in particular because massless higher-spin particles in flat space are known to be inconsistent with unitarity and locality.
This has been demonstrated in a number of ways \cite{Weinberg:1964ew,Coleman:1967ad,Aragone:1979hx,Weinberg:1980kq,Berends:1984rq,Porrati:2008rm,Porrati:2012rd,Henneaux:2012wg,Henneaux:2013gba,Hinterbichler:2017qcl}, including from a purely on-shell perspective \cite{Benincasa:2007xk,Schuster:2008nh,McGady:2013sga}.
Nevertheless, the simplicity of our result and of the way we derive it strongly motivates us to present it.
It is also worth noting that the massless higher-spin no-go theorems can be evaded in curved space \cite{Fradkin:1987ks,Vasiliev:1990en,Vasiliev:1999ba,Kogan:2000uy,Porrati:2000cp,Vasiliev:2003cph,Anninos:2011ui,Karch:2001jb}, of potential relevance in the context of inflation \cite{Arkani-Hamed:2015bza,Anninos:2019nib}. 
See also \cite{Bekaert:2010hw,Rahman:2012thy,Rahman:2015pzl} for reviews of higher-spin fields.

We do not attempt to construct a complete interacting theory of massive higher-spin particles in flat space, nor do we claim that the high-energy limit of a massive exchange automatically defines such a theory \cite{Cucchieri:1994tx,Arkani-Hamed:2002bjr,Porrati:2008gv,Porrati:2008ha,Caron-Huot:2016icg,Bellazzini:2017fep,Bonifacio:2018aon,Afkhami-Jeddi:2018apj,Bonifacio:2018vzv,Bellazzini:2019bzh,Bellazzini:2023nqj}.
Rather, our goal is to identify a simple on-shell probe of the discontinuity between massive and massless exchanges, independently of the existence of a consistent ultraviolet completion.

The paper is organised as follows. Section~\ref{sec:setup} sets up the on-shell three-point amplitude with two scalars and a massive spin-$S$ mediator and the resulting four-point amplitude. Section \ref{sec:residue-discontinuity} contains our main result: the Chebyshev--Legendre dichotomy of the $t$-channel residue and its interpretation as a higher-spin generalisation of the vDVZ discontinuity. Section~\ref{sec:nr-potential} translates the residue diagnostic into the imprint on the non-relativistic two-body potential. We conclude in Section~\ref{sec:conclusion}. Spinor-helicity conventions, the case-by-case high-energy expansion of the three-point amplitude, and a covariant rewriting in terms of source--source amplitudes that reproduces the standard vDVZ structures are collected in the appendices.

\section{Massive Higher-Spin Mediators}\label{sec:setup}

\subsection{Three-Point Amplitude}
We are interested in the three-point interaction between two massive scalars and a massive spin-$S$ particle, the basic building block for scalar scattering via higher-spin exchange. We denote the mediator mass by $m$ and the scalar masses by $M$. 
For two scalars of the \emph{same} species,
\begin{equation}
\amp_3(\mathbf{1}_{\phi},\mathbf{2}_{\phi},\mathbf{3}^{S})^{\{I_1\ldots I_{2S}\}},
\label{eq:3pt_massive}
\end{equation}
where $\{I_1,\ldots,I_{2S}\}$ are the $SU(2)$ little-group indices of the massive spin-$S$ leg, which we will suppress via the bold notation of \cite{Arkani-Hamed:2017jhn}. The little-group covariant on-shell analysis (details are provided in Appendix~\ref{app:detailed_scalar_scatering}) reduces to a single Lorentz structure, corresponding to minimal coupling, which admits the equivalent compact forms
\begin{equation}
\amp_3(\mathbf{1}_{\phi},\mathbf{2}_{\phi},\mathbf{3}^{S})
=
\frac{g}{\Lambda^{3S-1}}(-1)^S
\big\langle \mathbf{3}\big|
p_1 \bar{p}_2
\big|\mathbf{3}\big\rangle^{S}
=
\frac{g}{\Lambda^{3S-1}}\left(\frac{m}{2}\right)^{S}
\big\langle \mathbf{3}\big|
(p_1 - p_2)
\big|\mathbf{3}\big]^{S},
\label{eq:3pt_scalar_scalar_spinS}
\end{equation}
where, for convenience, we have defined $g$ as a (real) dimensionless coupling constant and encoded the mass dimension of the interaction via suitable powers of a scale $\Lambda$. The two forms are related by momentum conservation and the massive Dirac equation; we use them interchangeably. A useful polarisation-tensor form,
\begin{equation}
\amp_3(\mathbf{1}_{\phi},\mathbf{2}_{\phi},\mathbf{3}^{S})
=
\frac{g}{\Lambda^{3S-1}}
\left(\frac{m^2}{\sqrt{2}}\right)^{S}
(p_1-p_2)^{\mu_1}\cdots(p_1-p_2)^{\mu_S}\,
\bm{\varepsilon}_{\mu_1\ldots\mu_S}(p_3),
\label{eq:3pt_scalar_scalar_spinS_pol}
\end{equation}
is derived in Appendix~\ref{app:polarisations}. Under interchange of the two scalar legs, $1\leftrightarrow 2$, the amplitude picks up a factor $(-1)^S$. For identical scalars, Bose symmetry therefore enforces $g=0$ for odd $S$. A non-vanishing odd-$S$ coupling requires distinguishable legs (e.g., by charge or flavour), i.e., $\amp_3(\mathbf{1}_{\phi_i},\mathbf{2}_{\phi_j},\mathbf{3}^{S})$ with $i \neq j$. For odd $S$, therefore, it is implicitly understood that we consider scattering of complex scalars, with $\amp_3(\bm{1}_{\phi},\bm{2}_{\phi^*},\bm{3}^{S})$, instead of \eqref{eq:3pt_massive}, given by \eqref{eq:3pt_scalar_scalar_spinS}.

\subsection{Four-Point Amplitude Residue}

The tree-level four-point amplitude $\amp(\phi_1\phi_2\to\phi_1\phi_2) \equiv \amp_4^{(m)}$, up to contact terms, follows from factorisation on the massive
spin-$S$ pole. For odd $S$ we consider scattering of complex scalars, where $\phi_1$ and $\phi_2$ have the same charge. Writing $q=-(p_1+p_3)$ and $t=q^2$, the $t$-channel contribution takes the standard gluing form 
\begin{equation}
\lim_{t\to m^2} (t-m^2)\, \amp_4^{(m)}
=
-\Big[
\amp_3(\bm{1}_{\phi_1},\bm{3}_{\phi_1},\bm{q}^{S})^{\{I\}}
\,\epsilon\cdot\ldots\cdot \epsilon\,
\amp_3(\bm{2}_{\phi_2},\bm{4}_{\phi_2},-\bm{q}^{S})^{\{J\}}
\Big]_{q^2=m^2},
\label{eq:4pt_gluing}
\end{equation}
with the $\epsilon$-tensors contracting the $2S$ little-group indices of the internal massive state; see Appendix~\ref{app:onshell:gluing} for details.%
\footnote{For odd \(S\), due to the charge flow, the residue will be given by 
\begin{equation*}
\lim_{t\to m^2} (t-m^2)\, \amp_4^{(m)}
=
-\Big[
\amp_3(\bm{1}_{\phi_1},\bm{3}_{\phi^*_1},\bm{q}^{S})^{\{I\}}
\,\epsilon\cdot\ldots\cdot \epsilon\,
\amp_3(\bm{2}_{\phi_2},\bm{4}_{\phi_2^*},-\bm{q}^{S})^{\{J\}}
\Big]_{q^2=m^2}.
\label{eq:4pt_gluing_odd}
\end{equation*}}
In $D=4$, the residue reorganises into the expected Legendre angular dependence. Indeed, defining
\begin{equation}
z \equiv \frac{2s+m^2-2M_1^2-2M_2^2}{\sqrt{(m^2-4M_1^2)(m^2-4M_2^2)}},
\label{eq:z_def}
\end{equation}
with $s = (p_1+p_2)^2$, the residue of the four-point amplitude from exchange of a single massive spin-$S$ state
in the $t$-channel is
\begin{equation}
\lim_{t\to m^2} (t-m^2)\,\amp_4^{(m)}
=
(-1)^{S+1} \left(\frac{g\,m^{2S}}{\Lambda^{3S-1}}\right)^2
\frac{(S!)^2}{(2S)!}\,
\big((m^2-4M_1^2)(m^2-4M_2^2)\big)^{S/2}\,
P_S(z),
\label{eq:4pt_general_spinS}
\end{equation}
in agreement with the well-known
low-spin cases $S=0,1,2$. We detail the calculation of the combinatorial prefactors in Appendix~\ref{app:detailed_scalar_scatering}.
The Legendre polynomial $P_S(z)$ is the residue's only spin-dependent kinematic content.
Note that \eqref{eq:4pt_general_spinS} is nothing but the spin-$S$ contribution to the $t$-channel partial-wave decomposition of the amplitude. The corresponding angular variable, $\cos \theta_t$, is the cosine of the scattering angle in the frame where $\vec{p}_1$ and $\vec{p}_3$ are collinear, or, equivalently, the scattering angle of the crossed process $\phi_1\phi^*_1 \to \phi_2 \phi^*_2$ in its center-of-mass frame. \eqref{eq:z_def} gives the covariant expression of $\cos \theta_t$, analytically continued to the physical $s$-channel region and evaluated at $t = m^2$, i.e., $\cos\theta_t|_{t=m^2} = z$; see also \eqref{eq:z_def_source_source}.

\section{The Chebyshev--Legendre Dichotomy at the
Pole}\label{sec:residue-discontinuity}

We now compare the strictly massless spin-$S$ exchange with the massless limit, $m/M_i\to0$, of the corresponding massive spin-$S$ exchange. Since the center-of-mass energy satisfies $E \geqslant M_i$, we also have $m/E\to0$ in this limit.

\subsection{Strictly Massless Exchange}

We begin the comparison by determining the relevant massless spin-\(S\) pole of the four-point amplitude in a similar way to that for the massive exchange. The relevant
three-point amplitude with two equal-mass scalar legs and one massless leg of helicity \(h\) is written in the usual \(x\)-factor form
\begin{equation}
\amp_3(\bm 1_\phi,\bm 2_\phi,3^h)=\frac{\gm M^S}{\Mm^{S-1}}\,x_{12}^h,
\label{eq:3pt_massless}
\end{equation}
where $S = |h|$ and we have chosen $\gm$ to be real and the normalization to match that of a massless graviton, $h = \pm 2$, with $\Mm/\gm = \Mp$. 
For identical scalars of mass $M$,
\begin{equation}
x_{12}=
\frac{\langle\zeta|p_1|3]}{M\langle\zeta3\rangle}
=
\frac{\langle\zeta|(p_1-p_2)|3]}{2M\langle\zeta3\rangle},
\end{equation}
where the reference spinor \(\zeta\) drops out of gauge-invariant amplitudes. 
Massless gluing then yields for the pole of the four-point amplitude\begin{equation}
\begin{split}
\lim_{t\to 0} t\,\amp_4^{(0)}
&=
-\left[\sum_{h=\pm S}
\amp_3(\bm 1_{\phi_1},\bm 3_{\phi_1},q^h)\,
\amp_3(\bm 2_{\phi_2},\bm 4_{\phi_2},-q^{-h})\right]_{q^2=0}
\\
&=
(-1)^{S+1}
\frac{\gm^2 (M_1M_2)^S}{\Mm^{2(S-1)}}
\left(\frac{x_{13}^{S}}{x_{24}^{S}}+\frac{x_{24}^{S}}{x_{13}^{S}}\right)_{q^2=0},
\end{split}
\label{eq:massless_exchange_four_point}
\end{equation}
where the $(0)$ superscript indicates that the four-point amplitude $\amp(\phi_1\phi_2\to\phi_1\phi_2)$ is mediated by a massless particle.%
\footnote{We note that our convention for the analytic continuation of the spinors is given by $[-q| = + [q|$, $\langle-q| = - \langle q|$; this implies, in particular, $\amp_3(\bm 2_{\phi_2},\bm 4_{\phi_2},-q^{-h}) = (-1)^S (\gm M_2^S/\Mm^{S-1})\,x_{24}^{-h}$.}
We define
\(y \equiv \frac{1}{2}(x_{13}/x_{24}+x_{24}/x_{13})|_{t = 0}\),
which evaluates to
\begin{equation}
y=\frac{s-M_1^2-M_2^2}{2M_1M_2},
\label{eq:y_mandelstam}
\end{equation}
i.e., the relative Lorentz factor of the two
scalars, $y=p_1\cdot p_2/(M_1M_2)$. 
The helicity sum in \eqref{eq:massless_exchange_four_point} produces the characteristic combination $x^S+x^{-S}$, which is the defining identity for the Chebyshev polynomial of the first kind:
\begin{equation}
x^S+\frac{1}{x^S}
=
2\,T_S\!\left(\frac{x+x^{-1}}{2}\right)
=
2\,T_S\!\left(y\right)
\label{eq:chebyshev_from_x_factor}
\end{equation}
Hence, the residue of the strictly massless exchange is
\begin{equation}
\lim_{t\to 0} t\,\amp_4^{(0)}
=(-1)^{S+1}\sigma_0
\frac{\gm^2 (M_1M_2)^S}{\Mm^{2(S-1)}}\,
2\,T_S\!\left(y\right),
\label{eq:massless_exchange_chebyshev}
\end{equation}
where $\sigma_0 = 1/2$ for $S = 0$ and $\sigma_0 = 1$ otherwise. Defining, for notational convenience, the $t$-channel massless residue
\begin{equation}
R_S^{(0)}(s) \equiv \lim_{t\to 0} t\,\amp_4^{(0)},
\end{equation}
we find for the lowest spins,
\begin{align}
R_1^{(0)}(s) &=  2\,\gm^2 M_1M_2 \,y, 
\label{eq:mlessR1}\\
R_2^{(0)}(s) &= -4\,\frac{\gm^2 (M_1M_2)^2}{\Mm^2}  \,\bigg(y^2-\frac{1}{2}\bigg),
\label{eq:mlessR2}\\
R_3^{(0)}(s)    &= 8\, \frac{\gm^2 (M_1M_2)^3}{\Mm^{4}} \bigg(y^3-\frac{3}{4}y\bigg).
\label{eq:mlessR3}
\end{align}
As will become clear in Section \ref{sec:nr-potential}, the residue is negative for even spins, being associated with an attractive force, while it is positive for odd spins, as it corresponds to a repulsive force between same-charge particles.

\subsection{The Massless Limit of Massive Exchange}

We compare the massless residue computed above to the massless limit of \eqref{eq:4pt_general_spinS}.
A standard analysis of the three-point amplitude (Appendix~\ref{app:he_limit_3pt}) shows that demanding a smooth, non-trivial transverse limit fixes the parametric scaling $\Lambda^{3S-1}\sim m^{2S}$. We are assuming this scaling from now on, to ensure a well-behaved yet non-vanishing massless limit of the scattering amplitude. We thus define
\begin{equation}
\Lambda^{3S-1} \equiv m^{2S}\Lambda'^{\,S-1}.
\label{eq:Lambda_scaling_generalS}
\end{equation}
For $S=2$, with $\Lambda' = \Mp$, $\Lambda$ coincides with the familiar Fierz-Pauli strong-coupling scale $\Lambda_5=(m^4M_{\rm Pl})^{1/5}$ \cite{Vainshtein:1972sx,Arkani-Hamed:2002bjr,Aubert:2003je,Cheung:2016yqr}.
Defining
\begin{equation}
R_S^{(m)}(s,m^2) \equiv \lim_{t\to m^2} (t-m^2)\,\amp_4^{(m)},
\end{equation}
and using \eqref{eq:Lambda_scaling_generalS} and the massless limit $m \ll M_i$ of \eqref{eq:z_def}
\begin{equation}
\lim_{m\to 0} z = y
\label{eq:z_to_y_over_2}
\end{equation}
we find that \eqref{eq:4pt_general_spinS} yields 
\begin{equation}
\lim_{m \to 0} R_S^{(m)}(s,m^2) =
(-1)^{S+1}
\frac{4^S(S!)^2}{(2S)!}\,
\frac{g^2 (M_1M_2)^S}{\Lambda'^{\,2(S-1)}}
P_S\!\left(y\right).
\label{eq:he_limit_massive_exchange}
\end{equation}
For the lowest spins,
\begin{align}
\lim_{m \to 0} R_1^{(m)}(s,m^2) &=  2\,g^2 M_1M_2 \,y,
\label{eq:m0R1}\\
\lim_{m \to 0} R_2^{(m)}(s,m^2) &= -4\,\frac{(g M_1M_2)^2}{\Lambda'^2}  \,\left(y^2-\frac{1}{3}\right),
\label{eq:m0R2}\\
\lim_{m \to 0} R_3^{(m)}(s,m^2) &=  8\,\frac{g^2 (M_1M_2)^3}{\Lambda'^{4}} \,\left(y^3-\frac{3}{5}y\right).
\label{eq:m0R3}
\end{align}
For $S=1$, the functional form of the residue agrees with the strictly massless result, whereas from $S=2$ onward, there is a mismatch.

\subsection{Chebyshev vs Legendre: Polynomial Measure of the Discontinuity}

Direct inspection of Eqs.~\eqref{eq:mlessR1}--\eqref{eq:mlessR3} and \eqref{eq:m0R1}--\eqref{eq:m0R3}, shows that for $\gm = g$ and $\Mm = \Lambda'$, the highest-power terms, $y^S$, of the massless residue and the high-energy limit of the massive residue match. This holds at any spin, since the coefficients of the respective highest-degree terms are $T_S(x)|_{x^S} = 2^{S-1}$ and $P_S(x)|_{x^S} = 2^{-S}(2S)!/(S!)^2$, compensating the prefactors in \eqref{eq:massless_exchange_chebyshev} and \eqref{eq:he_limit_massive_exchange}.
Fixing $\gm = g$ and $\Mm = \Lambda'$, one can then identify
\begin{equation}
\Delta_S(y) \equiv 
\frac{\Lambda'^{2(S-1)}}{(-1)^{S+1} g^2 (M_1M_2)^S}\left( R_S^{(0)} - \lim_{m \to 0} R_S^{(m)}\right)
= 2T_S(y)-\kappa_S P_S(y), \qquad \forall \, S \geq 1,
\label{eq:delta_def}
\end{equation}
with $\kappa_S = 4^{S}(S!)^2/(2S)!$, as a polynomial measure of the discontinuity.

The function $\Delta_S(y)$ starts at order $y^{S-2}$ for $S\ge2$, and only powers $y^{S-2k}$ appear. Explicitly,
\begin{align}
\Delta_1(y)&=0, \\
\Delta_2(y)&=-\frac{2}{3}, 
\label{eq:delta_examples}\\
\Delta_3(y)&=-\frac{6}{5}y. \\
 &\;\vdots \nonumber
\end{align}
The case $S=1$ is a useful sanity check. The $m\to0$ limit of the massive spin-1 exchange reproduces the strictly massless result, already at the level of the residue, consistent with the absence of a vDVZ-like discontinuity for QED and Yang--Mills.
For $S=2$, the strictly massless residue gives $T_2(y)=2y^2-1$, while the $m\to0$ limit of the massive residue gives $P_2(y)=\tfrac12(3y^2-1)$.
Matching the leading $y^2$ behaviour fixes $\kappa_2=8/3$, so that we obtain \eqref{eq:delta_examples}.
After matching the tensor $h=\pm2$ contribution, the remaining difference is a non-vanishing constant. 

This is the on-shell imprint of the vDVZ discontinuity: in the limit $m\to0$, the massive spin-$2$ exchange retains a finite scalar-like contribution and therefore does not reduce to strictly massless graviton exchange \cite{vanDam:1970vg,Zakharov:1970cc}.
For generic $S$, on the strictly massless side, the gluing in \eqref{eq:massless_exchange_four_point} sums only over the
two helicities $h=\pm S$, producing the Chebyshev structure $T_S(y)$ in \eqref{eq:massless_exchange_chebyshev}. 
On the massive side, the residue implicitly sums over the full spin-\(S\) multiplet. In the limit \(m\to0\), the additional helicity components do not automatically decouple; they survive with fixed relative weights determined by the massive representation and yield the Legendre structure $P_S(y)$ in \eqref{eq:he_limit_massive_exchange}.%
\footnote{The parity selection of these helicities is already encoded in
\[
P_S(-y)=(-1)^S P_S(y),
\]
so only monomials \(y^{S-2k}\) appear. In the high-energy limit, the corresponding polarisation decomposition is
\[
\bm\varepsilon_{\mu_1\cdots\mu_S}
\sim
(\varepsilon^++\varepsilon^-)_{\mu_1}\cdots
(\varepsilon^++\varepsilon^-)_{\mu_S},
\]
with longitudinal contributions suppressed as \(\varepsilon^{(0)} \to 
O(m/M)
\) relative to the transverse ones, since we are contracting them with a conserved current. Thus, the leading components have \(h=n_+-n_-=S-2k\), i.e., the same parity as \(S\). Note that these statements refer to the little-group frame adapted to the massless momentum, see Section \ref{app:polarisations:HE}}.
The Chebyshev--Legendre mismatch is a residue-level diagnostic of the discontinuity between a strictly massless exchange and the massless limit of a massive one.
This result agrees with the involved Lagrangian-based field-theoretic current-exchange calculations in \cite{Francia:2007qt,Francia:2008hd}, see in particular the note added in \cite{Francia:2007qt} and the
review \cite{Sagnotti:2010jt}, where the vDVZ discontinuity for generic $S$ is implicitly given through the explicit dependence of the current exchange on the space-time dimensionality; see also Appendix~\ref{app:source-source}.
Let us also note that the presence of a discontinuity for $S > 1$ was first noticed in \cite{Berends:1979je}, where it was also shown that, unless the
higher-spin source contains only its highest-spin component, or it is projected onto it  (i.e.~is transverse-traceless), the non-pole part of the amplitude diverges in the massless limit. Since this part is a contact term, it is not fixed by factorisation and lies outside the residue comparison performed here. 
The vDVZ discontinuity in massive (super)gravity has been previously derived with on-shell methods in \cite{Moynihan:2017tva,Burger:2020kle}.

Another, more physical  distinction between the massless residue and the massive one at high energies comes from matching the overall coupling normalisation (see also Section \ref{sec:nr-potential}). 
The comparison reduces then to the replacement of angular structures
\begin{equation}
T_S\!\left(y\right)
\quad\longleftrightarrow\quad
P_S\!\left(y\right).
\label{eq:T_vs_P}
\end{equation}
More precisely, the equality of the static forces fixes one overall coupling constant by matching the value of the residues at the forward point $y = 1$, where $T_S(1)=P_S(1)=1$,
\begin{equation}
g = \gm \left(\frac{\Lambda'}{\Mm}\right)^{S-1} \frac{\sqrt{2(2S)!}}{2^{S}S!} \qquad \text{for} \; S\geq 1.
\label{eq:gmatch}
\end{equation}
The remaining difference is therefore not an overall normalisation effect, but a difference in the polynomial dependence on the on-shell variable \(y\).

Before moving to the next section, let us stress that the analysis carried out here is a comparison of the \emph{$t$-channel factorisation residues}, not of the full off-pole
amplitude at a generic scattering angle. 
This establishes that the two exchanges are already inequivalent at the level of pole data.
Extending the comparison away from the pole would require additional dynamical input, in particular, the off-pole exchange numerator and, in general, contact terms. This fact will be apparent in the next section.

\section{Non-Relativistic Long-Range Potential and  \texorpdfstring{$v^2$}{velocity-squared} Correction}
\label{sec:nr-potential}

The Chebyshev--Legendre mismatch is a statement about the residue of an exchange pole. It is therefore natural to ask how this on-shell diagnostic would appear
in a physical observable. A simple probe is the long-range classical potential between two heavy scalar sources  since, at leading order in the (three-point) coupling constant, this is given by the residue we have computed; see, e.g., \cite{Chung:2018kqs}. In the exact static limit, $y = 1$, the leading order potential is insensitive to the distinction between the massless and the $m \to 0$ massive residues after matching the respective couplings according to \eqref{eq:gmatch}.
The first robust long-distance difference, therefore, cannot appear in the leading non-relativistic \(1/r\) force. Instead, it is controlled by the first departure from the forward point and arises at relative order \(v^2\) in the non-relativistic expansion. In other words, the difference is a long-range velocity-dependent correction proportional to the difference of
the forward slopes
\(
T_S'(1)-P_S'(1).
\)
A complementary imprint of the same trace-sector mismatch, comparing non-relativistic and relativistic sources, is discussed in Appendix~\ref{app:rel_vs_non-rel}. 

\subsection{Near-Forward Kinematics and Matching of the Static Force}
\label{subsec:near_forward_static_force}

We now isolate the part of the non-relativistic interaction that is fixed unambiguously by the exchange pole.  We do not assume an off-pole completion of the numerator.  Instead, we work directly with the pole residue and use it only to determine the coefficient of the long-range \(1/r\) force.\footnote{Terms regular at the pole, or terms proportional to positive powers of the momentum transfer after the pole is removed, correspond to contact interactions and are not included.}

We work in the centre-of-mass frame with equal external scalar masses $M_1 = M_2 = M$, with incoming momentum $p_1 = (E,\vec{p})$ and outgoing $p_3 = (E,\vec{p'})$. The momentum transfer is $\vec{q} = \vec{p'} - \vec{p}$ and the average $\vec{P}=(\vec{p}+\vec{p'})/2$.
In the non-relativistic limit, $|\vec{q}| \sim |\vec{p}| \sim |\vec{p'}| \sim |\vec{P}| \sim M v$, where $v \ll 1$ is the relative velocity.
Using the standard leading Born matching for two heavy scalars (see, e.g., \cite{Cheung:2018wkq}), the long-range classical potential is obtained by Fourier transforming the pole part of the exchange amplitude. 
Massless exchange leads to
\begin{equation}
V_S^{(0)} = \frac{R_S^{(0)}}{4E^2}\frac{1}{4\pi r},
\label{eq:Vmless}
\end{equation}
with the massless residue given in \eqref{eq:massless_exchange_chebyshev}.
Massive exchange, instead, gives rise to the Yukawa potential,
\begin{equation}
V_S^{(m)} = \frac{R_S^{(m)}}{4E^2}\frac{e^{-mr}}{4\pi r},
\label{eq:Vm}
\end{equation}
with the massive residue given in \eqref{eq:4pt_general_spinS}.
Since we are interested in its massless limit, $m \ll 1/r$ and $m \ll |\vec{p}|$, the discontinuity at the pole translates into the following difference of potentials,
\begin{equation}
\Delta V_S(r,\vec{P}) \equiv V_S^{(0)}-\lim_{m\to0}V_S^{(m)} =
\frac{1}{4\pi r} \left( R_S^{(0)} - \lim_{m \to 0} R_S^{(m)} \right) \frac{1}{4E^2}
\label{eq:DeltaV_long}
\end{equation}
Note that we have expressed the difference as a function of the average momentum. Several comments in this regard are in order. First, before Fourier transforming, the residues are expressed in terms of $s = (p_1 + p_2)^2 = 4 (M^2 + \vec{p}^2)$ in the center-of-mass frame. The angular variable on which the Chebyshev and Legendre (for $m \to 0$) depend is then
\begin{equation}
y = 1 + \frac{2\vec{P}^{\,2}}{M^2} + \frac{\vec{q}^{\,2}}{2M^2}
\end{equation}
In \eqref{eq:Vmless} and \eqref{eq:Vm}, contributions from the term proportional to the exchanged momentum have been dropped, since they do not contribute to the long-distance potential.
No dependence on the off-pole transfer momentum is introduced.

Second, and most importantly,
since the leading static forces of the massless and massive exchange in the high-energy limit can be made to match by fixing the respective couplings, see \eqref{eq:gmatch},
the residue difference in \eqref{eq:DeltaV_long} is non-vanishing starting at the first post-Newtonian order, $O(v^2)$,
\begin{equation}
R_S^{(0)} - \lim_{m\to0} R_S^{(m)}(s)
=
(-1)^{S+1}
\,2\,\frac{\gm^2 M^{2S}}{\Mm^{2(S-1)}} \frac{2\vec{P}^{\,2}}{M^2}
\left[
T_S'(1)-P_S'(1)
\right]
+ O(v^4).
\label{eq:residue_difference_velocity}
\end{equation}
Using $T_S'(1)=S^2$ and $P_S'(1)=S(S+1)/2$
we find
\begin{equation}
\Delta V_S(r,\vec{P})
=
(-1)^{S+1}
\,\gm^2 \left(\frac{M}{\Mm}\right)^{2(S-1)}
\,\frac{\vec{P}^{\,2}}{M^2}\,\frac{S(S-1)}{8\pi r}
+ O(v^4).
\label{eq:DeltaV_long_evaluated}
\end{equation}
Therefore, we find that if the static forces mediated by a massless higher-spin and a light massive higher-spin agree, the first long-range difference appears at relative order \(v^2\), through the slope of the pole residue with respect to the on-shell variable $y$.  
The discontinuity vanishes for \(S=0\) and \(S=1\), reproducing the well-known absence of a vDVZ-type discontinuity for scalar and vector exchange. 
The first non-trivial discontinuity arises at spin two, which is the post-Newtonian counterpart of the familiar statement that, in massive gravity (barring Vainshtein screening \cite{Vainshtein:1972sx,Deffayet:2001uk,Porrati:2002cp}), matching the gravitational attraction experienced by a non-relativistic body such as the Moon leaves a mismatch for ultra-relativistic probes, such as the bending of light by the Sun.

The result \eqref{eq:DeltaV_long_evaluated} should be understood as a statement about the long-range part fixed by the exchange pole. It does not determine contact terms, derivative contact terms, or a unique velocity-dependent potential representative. Those depend on the completion of the amplitude away from the pole and on field-redefinition or canonical-transformation conventions. The coefficient of \(\vec{P}^2/r\) above is insensitive to those ambiguities.\footnote{Indeed, as shown in, e.g., \cite{Holstein:2008sx,Bern:2019crd}, a canonical coordinate transformation modifies the potential at $O(\vec{P}^2)$ by a term proportional to $\vec{P}^2 - (\vec{P}\cdot\hat{r})^2$, where $\hat{r} = \vec{r}/r$, without affecting (gauge independent) observables. However, the difference of potentials in \eqref{eq:DeltaV_long_evaluated} is insensitive to such a choice of coordinates.}

\subsection{Forward Slopes, Little-Group Data, and Higher Dimensions}

There are three useful points to extract from \eqref{eq:DeltaV_long_evaluated}.
First, the leading static potential is insensitive to the discontinuity once the couplings are matched at the forward point,
\begin{equation}
T_S(1)=P_S(1)=1.
\end{equation}
The first long-range difference, therefore, appears only at relative order
\(v^2\). Second, the difference vanishes for \(S=0,1\) and is non-zero for
\(S\geq2\), consistent with the absence of a vDVZ-like discontinuity for
scalar and gauge-boson exchange and with the first non-trivial case occurring
at spin two. Third, its coefficient \(\frac12 S(S-1)\)
is the difference between the normalised near-forward Casimir data of the
\(SO(2)\) helicity-\(S\) partial wave and the \(SO(3)\) spin-\(S\) partial
wave,
\[
\widehat C_2\bigl(SO(2),S\bigr)=S^2,
\qquad
\widehat C_2\bigl(SO(3),S\bigr)=\frac12S(S+1).
\]

This admits a natural generalisation to arbitrary spacetime dimensions. In
\(d\) spatial dimensions, the sphere of directions is
\[
S^{d-1}\simeq \frac{SO(d)}{SO(d-1)},
\]
and its \(SO(d-1)\)-invariant, or zonal, harmonics are the Gegenbauer polynomials
\(C_S^{((d-2)/2)}(z)\). They reduce to Legendre polynomials for \(d=3\), while
for \(d=2\) they give the Chebyshev polynomials through a singular
Gegenbauer limit.

To implement the matching of the leading static potential at the forward
point, define the normalised Gegenbauer polynomial by
\begin{equation}
\widehat C_S^{(\lambda)}(z)
\equiv
\frac{C_S^{(\lambda)}(z)}{C_S^{(\lambda)}(1)},
\qquad
\widehat C_S^{(\lambda)}(1)=1.
\end{equation}
For \(S\geq1\), both the numerator and denominator vanish linearly as
\(\lambda\to0\), but their ratio has a finite limit
\[
\lim_{\lambda\to0}\widehat C_S^{(\lambda)}(z)=T_S(z).
\]

\noindent For a normalised
Gegenbauer polynomial, one has
\begin{equation}
\left.
\partial_z\widehat C_S^{(\lambda)}(z)
\right|_{z=1}
=
\frac{S(S+2\lambda)}{2\lambda+1}.
\label{eq:gegenbauer_slope_casimir}
\end{equation}
The numerator \(S(S+2\lambda)\) is the quadratic Casimir of the spin-\(S\)
symmetric-traceless representation of \(SO(2\lambda+2)\). The forward slope,
therefore, measures the normalised Casimir data controlling the first angular
departure from forward scattering.

In \(D\) spacetime dimensions, the massive and strictly massless little groups
are \(SO(D-1)\) and \(SO(D-2)\), respectively. Their corresponding normalised
angular structures are \(\widehat C_S^{\left(\frac{D-3}{2}\right)}(z)\) and \(\widehat C_S^{\left(\frac{D-4}{2}\right)}(z)\). 
This construction is naturally defined for \(D\geq4\). 

In \(D=4\), the
massless Gegenbauer parameter vanishes, and the controlled Chebyshev limit
above applies. In \(D=3\), however, the strictly massless little group is \(SO(1)\): There is no continuous transverse rotation group and no ordinary helicity tower analogous to \(SO(2)\) in four dimensions. Equivalently, the transverse sphere degenerates to the discrete space \(S^0\), so the smooth zonal-harmonic and near-forward derivative construction no longer applies.
For \(D=2\), the massive little group is also \(SO(1)\), and the massive construction degenerates as well.

The general-dimensional analogue of the difference in long-range potentials \eqref{eq:DeltaV_long_evaluated} is, therefore,
\begin{align}
\Delta V_S^{(D)}
&\sim
\frac{1}{r^{D-3}}
\bigg(
\partial_z\widehat C_S^{\left(\frac{D-4}{2}\right)}(z)
-
\partial_z\widehat C_S^{\left(\frac{D-3}{2}\right)}(z)
\bigg)\Bigg|_{z=1}
\nonumber\\
&\sim
\frac{1}{r^{D-3}}
\frac{S(S-1)}{(D-2)(D-3)}.
\label{eq:general_D_casimir_difference}
\end{align}
The poles at \(D=2,3\) thus signal the degeneration of the corresponding little-group harmonic construction, rather than a physical divergence.

Finally, the radial behaviour changes with dimension because the Fourier transform of \(1/(\bm q^2+m^2)\) depends on the number of spatial dimensions. The massive Green's function is expressed in terms of modified Bessel functions, while its massless long-range limit behaves as
\[
Y_0^{(D-1)}(r)\sim \frac{1}{r^{D-3}}.
\]
The radial falloff is therefore dimension-dependent, whereas the coefficient of the discontinuity is determined by the difference between the normalised near-forward Casimir data of the two little groups.

\section{Conclusion}\label{sec:conclusion}

We have given a uniform on-shell formulation of the discontinuity
between scalar scattering mediated by a strictly massless spin-$S$
particle and the high energy $m/E\to 0$ limit of scattering mediated by a massive spin-$S$ particle. The natural arena is the $t$-channel factorisation residue, which is fixed by on-shell three-point amplitudes alone and therefore independent of choices of gauge, field variables, or contact terms.

The main result is that in four dimensions, the two residues are related by a polynomial replacement $T_S(y) \leftrightarrow P_S(y)$. The Chebyshev structure is the fingerprint of the strictly massless helicity sum, which retains only $h=\pm S$; the Legendre structure is the fingerprint of the massive helicity multiplet, in which all helicities of the same parity as $S$ survive the high-energy limit with fixed relative weights. The on-shell mismatch reproduces the vDVZ discontinuity for $S=2$ and provides an elegant, simple, and explicit higher-spin generalisation. The polynomial measure $\Delta_S$ defined in \eqref{eq:delta_def} quantifies the discontinuity at every spin.

The concrete imprint of the diagnostic in the non-relativistic two-body potential is that, after matching the leading static force, the first long-range difference appears at relative order $v^2$ and is proportional to $S(S-1)/2$ -- a Casimir comparison between the $SO(2)$ helicity label of strictly massless exchange and the $SO(3)$ spin of the massive multiplet. This result has been generalized to higher dimensions. A consistent covariant rewriting in terms of source--source amplitudes (Appendix~\ref{app:source-source}) reproduces the standard vDVZ trace structures for $S=2$ and exhibits the even-versus-odd pattern of surviving lower-helicity sectors for higher spin.

The simplicity of the on-shell formalism, as also exemplified in other works (see \cite{Chung:2018kqs,Arkani-Hamed:2019ymq,Guevara:2019fsj,Aoude:2020ygw,Falkowski:2020mjq,Falkowski:2020fsu,Durieux:2019eor,Balkin:2021dko,Csaki:2020inw} for particularly interesting instances), lends itself to further investigations along these lines. For instance, one could extend the results to more general sources with matter content beyond scalars, compare the forces mediated by light but massive higher-spin particles with experimental data (see, e.g., \cite{Cembranos:2017vgi}), or systematically investigate how higher spins might serve as tree-level UV completions of non-minimal interactions \cite{Camanho:2014apa,Ruhdorfer:2019qmk,Caron-Huot:2022ugt}.

\bigskip
\section*{Acknowledgments}
This work has been supported by the Collaborative Research Center SFB1258, the Munich Institute for Astro-, Particle and BioPhysics (MIAPbP), and by the Excellence Cluster ORIGINS, which is funded by the Deutsche Forschungsgemeinschaft (DFG, German Research Foundation) under Germany’s Excellence Strategy – EXC 2094 – 39078331. 
The research of DH is partially supported by the International Max Planck
Research School (IMPRS) on “Elementary Particle Physics”.
The work of JS has been supported by grants RYC-2020-028992-I, PID2022-142545NB-C22, and CNS2023-145069, funded
by MICIU/AEI/10.13039/501100011033 and by the European Union (ESF/EU, ERDF/EU,
and NextGenerationEU/PRTR), and by CSIC-20223AT023 and “IFT Centro de Excelencia Severo
Ochoa CEX2020-001007-S”. 

\newpage

\begin{appendices}
\addtocontents{toc}{\protect\setcounter{tocdepth}{1}}
\numberwithin{equation}{section}

\section{On-Shell Formalism}\label{app:onshell}

This appendix briefly summarises the spinor-helicity conventions and the on-shell construction of three-point amplitudes used throughout the paper. We largely follow \cite{Dreiner:2008tw,Durieux:2019eor,Witten:2003nn,Arkani-Hamed:2017jhn,Baratella:2021guc} and adopt a mostly-minus metric convention.

\subsection{Conventions}\label{app:onshell:conv}

We map a four-vector to a bispinor via
\begin{equation}
p_{\alpha\dot\alpha}=p_\mu\sigma^\mu_{\alpha\dot\alpha},\qquad
\bar p^{\dot\alpha\alpha}=p_\mu\bar\sigma^{\mu,\dot\alpha\alpha},
\qquad
\sigma^\mu=(\mathbf 1,\vec\sigma),\ \ \bar\sigma^\mu=(\mathbf 1,-\vec\sigma).
\label{eq:conv:sigma-map}
\end{equation}
Spinor indices are raised/lowered with $\epsilon_{\alpha\beta}$ and $\epsilon_{\dot\alpha\dot\beta}$
\begin{equation}
\lambda^\alpha=\epsilon^{\alpha\beta}\lambda_\beta,\qquad
\tilde\lambda^{\dot\alpha}=\epsilon^{\dot\alpha\dot\beta}\tilde\lambda_{\dot\beta},
\qquad
\epsilon^{21}=-\epsilon^{12}=\epsilon_{12}=-\epsilon_{21}=-1,
\label{eq:conv:epsilon}
\end{equation}
where we follow the conventions of \cite{Dreiner:2008tw}. Consequently, for a bispinor we have
\(\bar p^{\dot\alpha\alpha}
=\epsilon^{\alpha\beta}\epsilon^{\dot\alpha\dot\beta}\,p_{\beta\dot\beta}\), such that
\(\det(p_{\alpha\dot\alpha})=\tfrac12\,p_{\alpha\dot\alpha}\bar p^{\dot\alpha\alpha}=m^2\). 

For massless particles, 
$\det p= m^2 = 0$ and $p_{\alpha\dot\alpha}$ has rank one, hence
\(p_{\alpha\dot\alpha}=\lambda_\alpha\,\tilde\lambda_{\dot\alpha}\).
For complex momenta, $\lambda$ and $\tilde\lambda$ are independent, while for real momenta they are related by complex conjugation \cite{Witten:2003nn}. The massless little group is $U(1)$, acting as
\(\lambda_\alpha\to w^{-1}\lambda_\alpha\),
\(\tilde\lambda_{\dot\alpha}\to w\,\tilde\lambda_{\dot\alpha}\),
and an $n$-point amplitude transforms as
$M(\ldots,w^{-1}\lambda_i,w\tilde\lambda_i,\ldots)
=w^{2h_i}\,M(\ldots,\lambda_i,\tilde\lambda_i,\ldots)$
for a massless leg $i$ of helicity $h_i$.

For massive particles, 
$m\neq0$, $p_{\alpha\dot\alpha}$ has rank two and can be written as the sum of two independent rank one tensors
\(p_{\alpha\dot\alpha}=\lambda^I_\alpha\,\tilde\lambda_{I,\dot\alpha}\),
$I=1,2$, with $\bar p^{\dot\alpha\alpha}=-\tilde\lambda^{I,\dot\alpha}\lambda_I^{\alpha}$ and indices raised (lowered) with $\epsilon^{IJ}$ ($\epsilon_{IJ}$).
The massive little group is $SU(2)$ and acts as
\(\lambda^I_\alpha\to {W^I}_J\,\lambda^J_\alpha\),
\(\tilde\lambda_{I,\dot\alpha}\to {(W^{-1})_I}^{\ J}\,\tilde\lambda_{J,\dot\alpha}\),
$W\in SU(2)$,
leaving $p_{\alpha\dot\alpha}$ invariant \cite{Arkani-Hamed:2017jhn}. 
Massive spinors satisfy the Dirac equation
\begin{equation}
p_{\alpha\dot\alpha}\tilde\lambda^{I,\dot\alpha}=m\,\lambda^I_\alpha,\qquad
p_{\alpha\dot\alpha}\lambda^{I,\alpha}=-m\,\tilde\lambda^I_{\dot\alpha}.
\label{eq:conv:massive-eom}
\end{equation}
and $(\lambda^I_\alpha)^* = {\rm sign}(p_0) \tilde{\lambda}_{I,\dot{\alpha}}$, $(\lambda_{I,\alpha})^* = -{\rm sign}(p_0) \tilde{\lambda}^I_{\dot{\alpha}}$ for real momenta.

For a more compact notation, we make use of the bracket notation where $|\lambda\rangle\equiv \lambda_\alpha$, $\langle\lambda|\equiv\lambda^\alpha$, $|\tilde\lambda]\equiv \tilde\lambda^{\dot\alpha}$, $[\tilde\lambda|\equiv\tilde\lambda_{\dot\alpha}$,
and define
\(\langle ij\rangle\equiv \lambda_i^\alpha\lambda_{j,\alpha}=-\langle ji\rangle\),
\([ij]\equiv \tilde\lambda_{i,\dot\alpha}\tilde\lambda_j^{\dot\alpha}=-[ji]\), with similar definitions for massive spinors. 

\subsection{Three-Particle Amplitudes}\label{app:onshell:M3}

For a massive spin-$S$ particle of momentum $p$, little-group covariance requires the amplitude to transform as a totally symmetric rank-$2S$ tensor under $SU(2)$. 
Factoring out the massive spinors,
\begin{equation}
\mathcal M(\ldots,p,\ldots)^{\{I_1\ldots I_{2S}\}}
=
\lambda_p^{I_1,\alpha_1}\cdots \lambda_p^{I_{2S},\alpha_{2S}}\;
\mathcal M_{\{\alpha_1\ldots\alpha_{2S}\}},
\label{eq:3pt_expand_lambdas}
\end{equation}
the stripped tensor $\mathcal M_{\{\alpha_1\ldots\alpha_{2S}\}}$ is now totally symmetric in its \(SL(2,\mathbb C)\) spinor indices. Its precise form is fixed by the
available Lorentz-covariant spinor-helicity building blocks, which we summarise here for the relevant three-point amplitudes of our analysis, where we take all momenta to be incoming.

\subsubsection{Two massive legs, one massless leg}\label{app:onshell:M3:2m1l}
With a single massless leg $3$ (helicity $h$), the analysis splits according to whether $m_1\neq m_2$ or $m_1=m_2$ \cite{Arkani-Hamed:2017jhn}.
For equal masses, $\lambda_3$ and $(p_1\tilde\lambda_3)/m$ become collinear, and one introduces the $x$-factor 
\begin{equation}
x\,\lambda_{3,\alpha}=\frac{ p_{1,\alpha\dot\alpha}}{m}\,\tilde\lambda_3^{\dot\alpha},
\qquad
x=\frac{\langle\zeta| p_1|3]}{m\langle\zeta 3\rangle}
=\frac{\langle\zeta| (p_1-p_2)|3]}{2m\langle\zeta 3\rangle},
\label{eq:M3:x-factor}
\end{equation}
leading to the basis decomposition
\begin{equation}
\mathcal M^{h}_{\{\alpha\},\{\beta\}}
=\sum_{i=|S_1-S_2|}^{S_1+S_2}\frac{g_i}{\Lambda^{n_i}}\,
x^{h+i}\,
\big(\lambda_3^{2i}\epsilon^{S_1+S_2-i}\big)_{\{\alpha\},\{\beta\}}.
\end{equation}

For scalar external legs ($S_1=S_2=0$), this automatically collapses to the single structure \eqref{eq:3pt_massless} used in the main text. 
The inverse of the \(x\)-factor is given by
\begin{equation}
x^{-1}\,\tilde{\lambda}_{3}^{\dot{\alpha}}=\frac{ \bar{p}_{1}^{\dot\alpha \alpha}}{m}\,\lambda_{3,\alpha},
\qquad
x^{-1}=\frac{[\zeta| p_1|3\rangle}{m[\zeta 3]}=\frac{[\zeta| (p_1-p_2)|3\rangle}{2m[\zeta 3]}.
\label{eq:M3:x-factor-inverse}
\end{equation}
Let us also note that, while $x$ is defined on strictly complex kinematics and therefore carries no reality condition, in the residue of a four-point amplitude, $x$ and $x^* = -x^{-1}$ are found on opposite sides of the cut.

\subsubsection{Three massive legs}\label{app:onshell:M3:3m}
When all legs are massive, a convenient Lorentz-tensor basis is provided by $\epsilon_{\alpha\beta}$ and
\begin{equation}
\mathcal O_{\alpha\beta}\equiv
 p_{1,\{\alpha\dot\gamma}\, p_2^{\dot\gamma}{}_{\beta\}},
\label{eq:M3:O-def}
\end{equation}
(symmetric in $\alpha,\beta$). The stripped amplitude can be expanded as \cite{Arkani-Hamed:2017jhn}
\begin{equation}
\mathcal M_{\{\alpha\},\{\beta\},\{\gamma\}}
=\sum_{i=0}^{1}\sum_{\sigma_i}
\frac{g_{\sigma_i}}{\Lambda^{\,3(S_1+S_2+S_3)-1-2i}}\,
\big(\mathcal O^{S_1+S_2+S_3-i}\epsilon^i\big)^{\sigma_i}_{\{\alpha\},\{\beta\},\{\gamma\}},
\label{eq:M3:3m}
\end{equation}
where \(i\) counts the number of epsilon tensors and \(\sigma_i\) labels the distinct ways the indices \(\{\alpha\}\), \(\{\beta\}\), \(\{\gamma\}\) can be distributed among the tensors. The sum over numbers of epsilon tensors can be truncated at one since \(\mathcal{O}_{\alpha\beta}\mathcal{O}_{\gamma\delta}-\mathcal{O}_{\gamma\beta}\mathcal{O}_{\alpha\delta}\sim \epsilon_{\alpha \gamma}\epsilon_{\beta \delta}\) and higher numbers of epsilon tensors can be rewritten in terms of tensors \(\mathcal{O}_{\alpha \beta}\).

\subsection{Gluing for Four-Point Amplitudes}\label{app:onshell:gluing}

Tree-level four-point amplitudes are meromorphic in kinematic invariants with simple poles from internal propagators going on-shell.

\subsubsection{Massless internal lines}
With an internal massless line of momentum $p$, the residue is a sum over the two helicities,
\begin{equation}
\lim_{p^2\to 0}p^2\,\mathcal M_4(1,2,3,4)
=-\sum_{h=\pm S}
\Big[\mathcal M_3(1,2,p^h)\,\mathcal M_3(-p^{-h},3,4)\Big]_{p^2=0}.
\label{eq:glue:massless}
\end{equation}

\subsubsection{Massive internal lines}
For a massive internal particle of spin $S$, one sums over its $SU(2)$ little-group indices \cite{Arkani-Hamed:2017jhn},
\begin{equation}
\lim_{p^2\to m^2}(p^2-m^2)\,\mathcal{M}
=-\mathcal{M}_{L,\{I_1\ldots I_{2S}\}}\,
\epsilon^{I_1J_1}\cdots\epsilon^{I_{2S}J_{2S}}\,
\mathcal{M}_{R,\{J_1\ldots J_{2S}\}},
\label{eq:glue:massive}
\end{equation}
where the sign of the exchanged momentum on one side of the cut is taken to be outgoing. 
We analytically continue massive spinors to negative momenta as $\lambda_{-p} = - \lambda_p$, $\tilde{\lambda}_{-p} = \tilde{\lambda}_p$, in line with the massless convention in Section \ref{sec:residue-discontinuity}. 
Useful contractions following from this choice are
\(\lambda_{p,\alpha}^I\lambda_{-p,\beta}^J\epsilon_{IJ}
=-m\,\epsilon_{\alpha\beta}\),
\(\tilde\lambda_{p}^{I,\dot\alpha}\tilde\lambda_{-p}^{J,\dot\beta}\epsilon_{IJ}
=-m\,\epsilon^{\dot\alpha\dot\beta}\),
and $|p^I\rangle[-p_I|= p$, $|p^I]\langle -p_I|=\bar{p}$, which reduce to external momenta after using momentum conservation at the vertices. 
Note that some signs are different here than in the case of both spinors being incoming; in particular, \(\lambda_{p,\alpha}^I\lambda_{p,\beta}^J\epsilon_{IJ}
=m\,\epsilon_{\alpha\beta}\),
\(\tilde\lambda_{p}^{I,\dot\alpha}\tilde\lambda_{p}^{J,\dot\beta}\epsilon_{IJ}
=-m\,\epsilon^{\dot\alpha\dot\beta}\), and $|p^I\rangle[p_I|= p$, $|p^I]\langle p_I|=-\bar{p}$.

\subsection{Polarisation Tensors and Physical States}\label{app:polarisations} 

Although the on-shell formalism fixes amplitudes from little-group covariance and analyticity, explicit polarisation tensors facilitate matching to off-shell Lagrangian descriptions. We summarise here only what is used in the main text.

\subsubsection{Massless polarisations}\label{app:polarisations:massless}
For a massless momentum $p_{\alpha\dot\alpha}=\lambda_\alpha\tilde\lambda_{\dot\alpha}$, spin-1 polarisations are defined using reference spinors $(\zeta,\tilde\zeta)$ \cite{Witten:2003nn, Elvang:2013cua}, 
\begin{align}
\varepsilon_\mu^+(p;\zeta)
&=\frac{[\tilde\lambda|\bar\sigma_\mu|\zeta\rangle}{\sqrt{2}\,\langle\lambda\zeta\rangle},
&
\varepsilon_\mu^-(p;\tilde\zeta)
&=\frac{\langle\lambda|\sigma_\mu|\tilde\zeta]}{\sqrt{2}\,[\tilde\lambda\tilde\zeta]}.
\label{eq:pol:massless:eps-vec}
\end{align}
They satisfy transversality $p^\mu\varepsilon_\mu^\pm=0$ and under Lorentz transformations introduce the gauge redundancy $\varepsilon_\mu^\pm\to \varepsilon_\mu^\pm+\alpha\,p_\mu$. Higher-spin massless polarisations are obtained as symmetric products.

\subsubsection{Massive spin-$S$ polarisations}\label{app:polarisations:massive1}
A convenient massive spin-1 polarisation bi-vector is \cite{Chung:2018kqs}
\begin{equation}
\varepsilon^{IJ}_\mu(p)=
\frac{\langle p^{\{I}|\sigma_\mu| p^{J\}}]}{\sqrt{2}\,m},
\label{eq:pol:massive:epsIJ}
\end{equation}
transverse and covariant under $SU(2)$. With Clebsch--Gordan normalisation,
$\bm{\varepsilon}^{+}_\mu\equiv
\varepsilon^{11}_\mu$,
$\bm{\varepsilon}^{-}_\mu\equiv\varepsilon^{22}_\mu$, and
$\bm{\varepsilon}^{(0)}_\mu\equiv (\varepsilon^{12}_\mu+\varepsilon^{21}_\mu)/\sqrt2$
satisfy the standard orthonormality and completeness relations.
Higher-spin massive polarisations are built as symmetrised products:
\begin{equation}
\varepsilon^{I_1\ldots I_{2S}}_{\mu_1\ldots \mu_S}(p)=\frac{\langle p^{\{I_1}|\sigma_{\mu_1}| p^{I_2}]
\cdots
\langle p^{I_{2S-1}}|\sigma_{\mu_{S}}| p^{I_{2S}\}}]}
{(\sqrt{2}\,m)^{S}}.
\label{eq:pol:massive:spinS-generating}
\end{equation}
If $k$ of the $2S$ little-group indices take the value $2$, the corresponding helicity is $h=S-k$; the canonical normalisation is $\bm{\varepsilon}^{(S-k)}_{\mu_1\ldots\mu_S}
=
\sqrt{\binom{2S}{k}}\,\varepsilon^{1\ldots 1\,2\ldots 2}_{\mu_1\ldots\mu_S}$. 

\subsubsection{High-energy limit of massive polarisations}
\label{app:polarisations:HE}
The high-energy limit is taken at a fixed null direction of motion as \(E\gg m\), such that the massive \(SU(2)\) little group contracts to the massless \(ISO(2)\) little group. The translational part acts trivially for finite-helicity representations, leaving the \(U(1)\simeq SO(2)\) helicity subgroup. Accordingly, the massive spin-\(S\) multiplet decomposes into massless helicity components.

Massive spinors can be conveniently decomposed as \cite{Arkani-Hamed:2017jhn}
\begin{align}
\lambda_{\alpha}^I
&=
\lambda_{\alpha} \zeta_-^I
+
\eta_{\alpha} \zeta_+^I,
\\
\tilde\lambda_{\dot\alpha}^I
&=
\tilde\lambda_{\dot\alpha}\zeta_+^I
+
\tilde\eta_{\dot\alpha}\zeta_-^I.
\end{align}
where the vectors $\zeta_\pm^I$ constitute a basis for the little group, with $\zeta_+^I \epsilon_{IJ} \zeta_-^I = 1$.
Massive momenta therefore decompose as $p = |\lambda\rangle[\tilde\lambda|+|\eta\rangle[\tilde\eta|$, with $\langle \lambda \eta \rangle = [\tilde\lambda \tilde\eta] = m$ and the first term identified, in the center-of-mass frame $p^\mu = (E,0,0,p)$, with the large momentum eigenvalue, $E+p \simeq 2E$, and the second piece with the small one, $E-p \simeq m^2/2E$ at leading order in $m/E$.

The high-energy expansion corresponds to \cite{Arkani-Hamed:2017jhn,Chung:2018kqs,Durieux:2019eor}
\begin{equation}
\frac{\lambda_a\lambda_b}{\sqrt{m_am_b}} \gg 1, \quad [\tilde\eta_a \tilde\eta_b] \ll 1,
\end{equation}
for particles $a,b$.

At the level of polarisation vectors, 
the transverse polarisation reduce smoothly to their massless counterparts,
\begin{equation}
\bm{\varepsilon}^{\pm}_\mu(p)
\xrightarrow{E\gg m}
\varepsilon^\pm_\mu(k;q) + O\left(\frac{m}{E}\right),
\label{eq:pol:HE:transverse}
\end{equation}
where the momentum $k$ is massless and $q$ is a 
(null) reference vector.
The longitudinal polarisation instead behaves as
\begin{equation}
\bm{\varepsilon}^{(0)}_\mu(p)
\xrightarrow{E\gg m}
\frac{k_\mu}{m}
+
O\left(\frac{m}{E}\right).
\label{eq:pol:HE:longitudinal}
\end{equation} 
The leading 
\(k_\mu/m\) term is therefore not suppressed at the level of the polarisation vector itself.
However, for a conserved current,
\begin{equation}
k_\mu J^\mu=0,
\end{equation}
it decouples, and one finds, in the high-energy limit,
\begin{equation}
J^\mu\bm{\varepsilon}^{(0)}_\mu(p)
=
O\left(\frac{m}{E}\right).
\label{eq:pol:HE:longitudinal_current}
\end{equation}

In terms of the full \(SU(2)\)-covariant vector polarisation, this contains transverse components proportional to
\(
(\zeta_-^2)^{\{IJ\}}\) and 
\((\zeta_+^2)^{\{IJ\}}\), together with a mixed component proportional to
\(\zeta_+^{\{I}\zeta_-^{J\}}\).
The latter cannot be discarded as a tensor identity. 
However, after contraction with a conserved current, it is suppressed according to
\eqref{eq:pol:HE:longitudinal_current}, and hence
\begin{equation}
J^\mu\varepsilon^{IJ}_\mu(p)
\xrightarrow{E\gg m}
(\zeta_-^2)^{\{IJ\}}
J^\mu \varepsilon^-_\mu(k;q)
+
(\zeta_+^2)^{\{IJ\}}
J^\mu \varepsilon^+_\mu(k;q)
+
O\left(\frac{m}{E}\right).
\label{eq:pol:HE:SU2}
\end{equation}

For spin \(S\), consider the contraction with the conserved symmetric scalar current appearing in the factorisation residue,
\begin{equation}
p_{\mu_1}
J^{\mu_1\cdots\mu_S}=0.
\end{equation}
Every longitudinal insertion contains a leading factor \(p_{\mu_i}/m\), which vanishes upon contraction with this current, while its remaining contribution is suppressed by \(m/E\). Consequently, only products of transverse polarisations can contribute at leading order. 
If such a product contains \(n_+\) positive-helicity and \(n_-\) negative-helicity vectors, with \(n_++n_-=S\), its total helicity is
\begin{equation}
h=n_+-n_-=S-2n_-.
\end{equation}
Thus, the only helicity components that can remain unsuppressed in this conserved-current contraction are

\begin{equation}
h=S,S-2,\ldots,-S.
\label{eq:pol:HE:unsuppressed}
\end{equation}

This same-parity selection rule is therefore a statement about the factorisation residue obtained after contraction with the conserved scalar current, rather than an equality for the uncontracted massive polarisation tensor.

Several comments are in order. First, for the three-point amplitude between two massive scalars and a massive spin-$S$ particle, \eqref{eq:3pt_massive}, the counting $\eta_3, \tilde{\eta_3} = O(m)$ is relative to the scalar mass, i.e.~the proper suppression is $O(m/M)$: the three-point amplitude has no kinematic invariant besides $m$ and $M$, and its little-group components are invariant under
boosts along the mediator momentum.
Second, the suppression of the helicity components other than \eqref{eq:pol:HE:unsuppressed} holds in the little-group frame adapted to the massless momentum $k$, which in the rest frame of the mediator this is the frame whose quantisation axis is orthogonal to the scalar spatial momentum up to $O(m/M)$.
In a generic frame, instead, all the helicity components are of the same order.
Finally, the exact residue \eqref{eq:4pt_general_spinS} receives contributions from all $2S+1$ components in the rest frame of the mediator, and the parity property $P_S(-z) = (-1)^S P_S(z)$ follows from the addition theorem for spherical harmonics.

\section{Details of the Scalar--Scalar--Spin-\texorpdfstring{$S$}{S} Massive Amplitudes}
\label{app:detailed_scalar_scatering}

\subsection{Three-Point Amplitude}

The relevant massive scalar--scalar--spin-\(S\) stripped three-point amplitude can be obtained from equation \eqref{eq:M3:3m} setting \(S_1=S_2=0\), which yields 
\begin{equation}
\mathcal M_{3,\{\alpha_1\ldots\alpha_{2S}\}}
=\sum_{i=0}^{1} g_i\,\big(\mathcal{O}^{S-i}\epsilon^i\big)_{\{\alpha_1\ldots\alpha_{2S}\}}.
\end{equation}
The \(\epsilon\)-structure is eliminated when the stripped three-point amplitude is contracted with the spinors of the massive spin-\(S\) particles, since the little-group indices of the spin-\(S\) particle are totally symmetrised and thus any contraction that induces an antisymmetric pairing of two identical massive spinors vanishes (in particular
$\langle 3^{\{I} 3^{J\}}\rangle=0$
). Only the purely $\mathcal{O}^S$ term contributes, leading to the compact form \eqref{eq:3pt_scalar_scalar_spinS} of the main text.
The polarisation-tensor representation \eqref{eq:3pt_scalar_scalar_spinS_pol} follows by first using momentum conservation together with the equation of motion of the massive particles and writing $( p_1-p_2)_{\alpha\dot\alpha}=(p_1-p_2)^\mu\sigma_{\mu,\alpha\dot\alpha}$ and grouping factors into $\varepsilon_{\mu_1\ldots\mu_S}^{I_1\ldots I_{2S}}(p_3)$.

\subsection{Four-Point Amplitude}
\label{app:four_point_details}

For scalar external legs, the use of the stripped three-point amplitude is particularly convenient, since the internal little-group sum in \eqref{eq:4pt_gluing} is simply performed by the use of $\lambda_{q}^{I,\alpha} \epsilon_{IJ} \lambda_{-q}^{J,\beta} = m \epsilon^{\alpha \beta}$.
The residue, then, takes the compact form
\begin{equation}
\mathrm{Res}_{t}(\mathcal M_4^{S})
=
-\left(\frac{g}{\Lambda^{3S-1}}\right)^2
m^{2S}\,
(\mathcal{O}_{13}^{S})_{\{\alpha_1\ldots\alpha_{2S}\}}\,
(\mathcal{O}_{24}^{S})^{\{\alpha_1\ldots\alpha_{2S}\}}.
\label{eq:Res_t_compact}
\end{equation}
The resulting symmetrised contraction of spinor indices naively produces $(2S)!$ terms, but most are identical. Since $\mathcal{O}_{ij}^{\alpha\beta}$ is symmetric in its spinor indices, the fully symmetrized $(\mathcal{O}_{ij}^S)_{\{\alpha_1\dots\alpha_{2S}\}}$ is built from $S$ indistinguishable symmetric pairs, and the contraction of two such tensors reduces to the sum over all pairings (perfect matchings) of $2S$ indices, of which there are
\(
(2S-1)!! = (2S)!/(2^S\,S!)
\)
inequivalent ones. 
Evaluating the spinor traces in terms of Mandelstams produces the Legendre polynomial of \eqref{eq:4pt_general_spinS}, with $z$ given by \eqref{eq:z_def}.

\subsection{Detailed High-Energy Limit of the Three-Point Amplitude}\label{app:he_limit_3pt} 

This appendix records the explicit case-by-case high-energy analysis of the three-point amplitude. We take the scalar masses equal to $M$ and use the high-energy expansion of the massive spinors on leg~3 reviewed in Section \ref{app:polarisations:HE}. Throughout, terms involving the reference spinors $\eta_3,\tilde\eta_3$ are $O(m)$, i.e.~suppressed by $m/M$ relative to $\lambda_3,\tilde\lambda_3$ in the little-group frame adapted to the massless momentum (see the
remarks at the end of Section \ref{app:polarisations:HE}).

\paragraph{$S=1$.}
Starting from the $p_1\bar p_2$ representation of \eqref{eq:3pt_scalar_scalar_spinS} and using the high-energy decomposition on leg~3,
\begin{equation}
\begin{split}
\langle 3^{\{I}|  p_1 \bar{ p}_2|3^{J\}}\rangle
={}&(\zeta_-^2)^{\{IJ\}}
\langle 3| p_1 \bar{ p}_2|3\rangle
+(\zeta_+^2)^{\{IJ\}}
\langle \eta_3| p_1 \bar{ p}_2|\eta_3\rangle
\\
&+(\zeta_-\zeta_+)^{\{IJ\}}
\Big(\langle 3| p_1 \bar{ p}_2|\eta_3\rangle+\langle \eta_3| p_1 \bar{ p}_2|3\rangle\Big),
\end{split}
\label{eq:he_split_s1_p1p2}
\end{equation}
where the $\zeta_-^2$, $\zeta_+^2$, $\zeta_-\zeta_+$ components correspond to $h=-1,+1,0$ respectively. Using the massive Dirac equation ($p_3|3] =m|\eta_3\rangle$) to trade $\eta_3$ for $\lambda_3$ and momentum conservation,
\(\langle \eta_3| p_1 \bar{p}_2|\eta_3\rangle=[3|\bar{ p}_1  p_2|3]\),
while the $h=0$ piece necessarily retains at least one $\eta_3=O(m)$ and is thus subleading. Restoring the prefactor and using the high-energy (HE) reduction of polarisations \eqref{eq:pol:HE:transverse}--\eqref{eq:pol:HE:longitudinal},
\begin{equation}
\mathcal M_3(\bm 1_\phi,\bm 2_{\phi^*},\bm 3^{S=1})
\xrightarrow{\rm HE}
\frac{g}{\Lambda^2}\,\frac{m^2}{\sqrt{2}}\,(p_1-p_2)^\mu
\begin{cases}
\varepsilon_\mu^{-}, & h=-1,\\
O(m), & h=0,\\
\varepsilon_\mu^{+}, & h=+1.
\end{cases}
\label{eq:he_limit_s1_pol}
\end{equation}
A smooth, non-vanishing transverse limit requires $\Lambda\sim m$ for $S=1$. In the $x$-factor language, dropping the subleading $h=0$ piece,
\begin{equation}
\frac{g}{\Lambda^2}\frac{m}{2}\,\langle \bm 3|
(p_1-p_2)
|\bm 3]
\xrightarrow{\rm HE}
-\frac{g}{\Lambda^2}\,m^2M\left(
(\zeta_-^2)^{\{IJ\}}\frac{1}{x}+(\zeta_+^2)^{\{IJ\}}x
\right),
\label{eq:he_limit_s1_x}
\end{equation}
with $x$ as in \eqref{eq:M3:x-factor}.

\paragraph{$S=2$.}
Each massive spin-1 polarisation reduces to a sum of helicity $\pm1$ polarisations at leading order, so the massive spin-2 tensor contains, beyond $h=\pm2$, a leading $h=0$ component from the $\varepsilon^+\varepsilon^-+\varepsilon^-\varepsilon^+$ combination. Schematically,
\begin{equation}
\mathcal M_3(\bm 1_\phi,\bm 2_\phi,\bm 3^{S=2})
\xrightarrow{\rm HE}
\frac{g}{\Lambda^{5}}\left(\frac{\sqrt{2}m^2}{2}\right)^{2}(p_1-p_2)^\mu(p_1-p_2)^\nu
\begin{cases}
\varepsilon_\mu^-\varepsilon_\nu^-, & h=-2,\\
\varepsilon_\mu^+\varepsilon_\nu^-+\varepsilon_\mu^-\varepsilon_\nu^+, & h=0,\\
\varepsilon_\mu^+\varepsilon_\nu^+, & h=+2.
\end{cases}
\label{eq:he_limit_s2_pol}
\end{equation}
This is the on-shell manifestation of the vDVZ discontinuity at the level of the three-point amplitude. A natural parametrisation of the prefactor that keeps the transverse limit finite is $\Lambda^5\sim m^4\Lambda'$. 

\paragraph{General $S$.}
At leading order in the high-energy limit, the massive spin-$S$ polarisation behaves as
\begin{equation}
\bm\varepsilon_{\mu_1\ldots\mu_S}\ \sim\ \bigl(\varepsilon_{\mu_1}^{+}+\varepsilon_{\mu_1}^{-}\bigr)\cdots\bigl(\varepsilon_{\mu_S}^{+}+\varepsilon_{\mu_S}^{-}\bigr),
\label{eq:he_polarisation_binomial_app}
\end{equation}
so the surviving leading-order helicities have the same parity as $S$. The overall prefactor scales as
$g\,m^{2S}/\Lambda^{3S-1}$, and a finite transverse $m\to0$ limit fixes
\(\Lambda^{3S-1}\sim m^{2S}\,\Lambda'^{\,S-1}\),
generalising the $S=1,2$ analyses. This is the scaling used in \eqref{eq:Lambda_scaling_generalS}. 

For illustration, the $S=3$ case gives
\begin{equation}
\mathcal M_3(\bm 1_\phi,\bm 2_\phi,\bm 3^{S=3})
\xrightarrow{\rm HE}
-\frac{g}{\Lambda^{8}}\,m^{6}M^{3}
\left(\frac{1}{x^{3}}+\frac{3}{x}+3x+x^{3}\right),
\label{eq:he_limit_s3_x}
\end{equation}
making explicit both the survival of the $h=\pm 1$ modes and their relative weight, $3$ vs $1$, compared to the transverse $h=\pm 3$ modes. The corresponding four-point residue is exactly $P_3$ evaluated at $y$, as expected from the general result in the main text.

\section{Source--Source Amplitudes and the Helicity Content}
\label{app:source-source}

In this appendix, we give a covariant current-exchange cross-check of the Chebyshev--Legendre residue diagnostic. 
The main point is that the distinction between the strictly massless exchange and the \(m\to0\) limit of the massive exchange can be read directly from the trace weights of the corresponding projectors. 
The computation is deliberately performed with a simple probe current. This is enough for the residue diagnostic, which concerns the projector numerator on the pole, rather than the most general off-shell current or the full non-relativistic potential.

\subsection{Probe currents and trace sectors}

Consider a symmetric rank-\(S\) current, $J^{\mu_1\cdots\mu_S}$, coupled to a spin-\(S\) field. At the \(t\)-channel exchange pole, the amplitude factorises as
\begin{equation}
\mathcal A_S
\;\xrightarrow{t\to m^2}\;
\frac{1}{t-m^2}\,
J^{\mu_1\cdots\mu_S}
\Pi_{\mu_1\cdots\mu_S,\nu_1\cdots\nu_S}
J'^{\nu_1\cdots\nu_S},
\label{eq:source_source_factorisation}
\end{equation}
where \(\Pi\) is the relevant propagator numerator. Since we only want to probe the trace weights of the exchange projector \(\Pi\) to compare with our residue diagnostic, it is sufficient to use the minimal highest-derivative representative
\begin{equation}
J_{\mu_1\cdots\mu_S}=K_{\mu_1}\cdots K_{\mu_S},
\qquad
J'_{\nu_1\cdots\nu_S}=L_{\nu_1}\cdots L_{\nu_S}.
\label{eq:probe_current_def}
\end{equation}
Here \(K\) and \(L\) are the average scalar momenta at the two cubic vertices.
This current has the same highest-spin on-shell cubic structure as the usual minimal coupling and has non-vanishing traces on massive scalar kinematics. 
It is not meant to represent the most general off-shell current. 
Rather, it is a probe current for the propagator numerator: it cleanly exposes how the
projector weights the trace sectors. Improvements proportional to the exchange momentum are killed by transversality on the pole, while lower-derivative and
metric-trace terms change the detailed off-shell numerator and the potential, but not the universal trace weights of the projector. 
Its \(j\)-fold trace is schematically
\begin{equation}
J^{[j]}_{\mu_1\cdots\mu_{S-2j}}
\sim
(K^2)^jK_{\mu_1}\cdots K_{\mu_{S-2j}},
\label{eq:jfold_trace_probe_current}
\end{equation}
and hence
\begin{equation}
J^{[j]}\!\cdot J'^{[j]}
\sim
(K^2L^2)^j(K\!\cdot L)^{S-2j}.
\label{eq:jfold_trace_contraction}
\end{equation}

For external scalar masses \(M_1\) and \(M_2\), the exact kinematics on the massive exchange pole \(t=m^2\) are
\begin{equation}
K^2=M_1^2-\frac{m^2}{4},
\qquad
L^2=M_2^2-\frac{m^2}{4},
\qquad
K\cdot L
=
\frac{2s+m^2-2M_1^2-2M_2^2}{4}.
\label{eq:finite_mass_probe_kinematics}
\end{equation}
The normalised angular variable, as in \eqref{eq:z_def}, is
\begin{equation}
z
=
\frac{K\cdot L}{\sqrt{K^2L^2}}
=
\frac{2s+m^2-2M_1^2-2M_2^2}
{\sqrt{(m^2-4M_1^2)(m^2-4M_2^2)}}.
\label{eq:z_def_source_source}
\end{equation}
At finite \(m\), the contraction of the \(j\)-fold traces therefore takes the form
\begin{equation}
J^{[j]}\!\cdot J'^{[j]}
\sim
(K^2L^2)^{S/2}z^{S-2j}.
\label{eq:finite_mass_trace_sector_monomials}
\end{equation}

Since we want to compare the strictly massless residue against the zero-mass limit of the massive residue, we now take \(m\to0\) after evaluating the massive expression on its pole. 
In this limit,
\begin{equation}
K^2\longrightarrow M_1^2,
\qquad
L^2\longrightarrow M_2^2,
\qquad
z\longrightarrow
y\equiv
\frac{s-M_1^2-M_2^2}{2M_1M_2}.
\label{eq:y_limit_source_source}
\end{equation}
For equal external masses, \(M_1=M_2=M\), every trace sector consequently reduces to
\begin{equation}
J^{[j]}\!\cdot J'^{[j]}
\;\xrightarrow{m\to0}\;
M^{2S}y^{S-2j}.
\label{eq:trace_sector_monomials_appendix}
\end{equation}
Thus, after the zero-mass limit, the trace decomposition of the propagator numerator becomes a polynomial in the basis
\begin{equation}
y^S,\quad y^{S-2},\quad y^{S-4},\quad\ldots .
\end{equation}
The Chebyshev--Legendre diagnostic is precisely the statement that the strictly massless and the \(m\to0\) massive projectors assign different coefficients to this same trace basis.

\subsection{Massive spin-\texorpdfstring{\(S\)}{S} exchange and Legendre weights} 

For a massive spin-\(S\) boson in \(D\) dimensions, the propagator numerator can be written as a transverse traceless projector. We use the form of
\cite{Chandrasekaran_2018} (yet with mostly-minus metric signature),
\begin{align}
\Pi_{\mu_1\cdots\mu_S}^{\nu_1\cdots\nu_S}
&=
\sum_{j=0}^{\lfloor S/2\rfloor}
\left(-\frac12\right)^j
\frac{S!}{j!(S-2j)!}
\frac{(2S-2j+D-5)!!}{(2S+D-5)!!}
\nonumber\\
&\hspace{1.5cm}\times
\Pi_{(\mu_1\mu_2}\cdots
\Pi_{\mu_{2j-1}\mu_{2j}}
\Pi^{(\nu_1\nu_2}\cdots
\Pi^{\nu_{2j-1}\nu_{2j}}
\Pi_{\mu_{2j+1}}^{\nu_{2j+1}}\cdots
\Pi_{\mu_S)}^{\nu_S)} ,
\label{eq:massive_projector_chandrasekaran}
\end{align}
where \(\Pi_{\mu\nu}\) is the transverse spin-one projector. Contracted with conserved currents, this becomes a sum over trace sectors. 
In \(D=4\), the coefficient of the \(j\)-fold trace sector reduces to
\begin{equation}
(-1)^j
\frac{(2S-2j)!}{2^S j!(S-j)!(S-2j)!} \frac{2^S (S!)^2}{(2S)!}.
\label{eq:massive_trace_coeff_d4}
\end{equation}
Therefore, using the probe current \eqref{eq:probe_current_def}, we obtain
\begin{equation}
J\!\cdot\!\Pi^{(m)}_S\!\cdot\!J'
\;\propto\;
 \frac{2^S(S!)^2}{(2S)!} \;M^{2S}
\sum_{j=0}^{\lfloor S/2\rfloor}
(-1)^j
\frac{(2S-2j)!}{2^S j!(S-j)!(S-2j)!}
z^{S-2j}.
\label{eq:massive_source_sum}
\end{equation}
This is exactly the standard monomial expansion of the Legendre polynomial,
\begin{equation}
P_S(z)
=
\sum_{j=0}^{\lfloor S/2\rfloor}
(-1)^j
\frac{(2S-2j)!}{2^S j!(S-j)!(S-2j)!}
z^{S-2j}.
\label{eq:legendre_monomial_definition}
\end{equation}
Hence
\begin{equation}
J\!\cdot\!\Pi^{(m)}_S\!\cdot\!J'
\;\propto\;
\frac{2^S(S!)^2}{(2S)!}\; M^{2S}\;P_S(z).
\label{eq:massive_projector_legendre_appendix}
\end{equation}
Consequently, the \(m\to0\) limit of the massive spin-\(S\) projector acts on the trace basis with Legendre weights.

\subsection{Strictly massless exchange and Chebyshev weights}

The strictly massless current exchange is governed by the transverse traceless projector appropriate to massless Fronsdal fields. 
Following
\cite{Francia:2007qt}, the contraction with conserved currents can be written as
\begin{equation}
J\!\cdot\!\Pi^{(0)}_S\!\cdot\!J'
=
\sum_{n=0}^{\lfloor S/2\rfloor}
\rho_n(D-2,S)\,
\frac{S!}{n!(S-2n)!2^n}
J^{[n]}\!\cdot J'^{[n]},
\label{eq:francia_massless_current_exchange}
\end{equation}
with coefficients that obey the recursion
\begin{equation}
\rho_{n+1}(D,S)
=
-\frac{\rho_n(D,S)}{D+2(S-n-2)},
\qquad
\rho_0(D,S)=1.
\label{eq:rho_recursion_general}
\end{equation}
Solving this recursion in $D=4$, where the relevant argument is \(D-2=2\), 
gives
\begin{equation}
\rho_n(2,S)
=
(-1)^n\frac{(S-n-1)!}{2^n(S-1)!}.
\label{eq:rho_solution_d4}
\end{equation}
Substituting \eqref{eq:rho_solution_d4} into
\eqref{eq:francia_massless_current_exchange}, gives
\begin{equation}
J\!\cdot\!\Pi^{(0)}_S\!\cdot\!J'
=
\sum_{n=0}^{\lfloor S/2\rfloor}
(-1)^n
\frac{S(S-n-1)!}{n!(S-2n)!}
\frac{1}{2^{2n}}
J^{[n]}\!\cdot J'^{[n]} .
\label{eq:massless_trace_coeff_d4}
\end{equation}
Using again the probe current \eqref{eq:probe_current_def}, this becomes
\begin{equation}
J\!\cdot\!\Pi^{(0)}_S\!\cdot\!J'
\;\propto\;
M^{2S}
\sum_{n=0}^{\lfloor S/2\rfloor}
(-1)^n
\frac{S(S-n-1)!}{n!(S-2n)!}
\frac{1}{2^{2n}}
z^{S-2n}.
\label{eq:massless_source_sum}
\end{equation}
The Chebyshev polynomial admits the expansion
\begin{equation}
T_S(z)
=
\frac{S}{2}
\sum_{n=0}^{\lfloor S/2\rfloor}
(-1)^n
\frac{(S-n-1)!}{n!(S-2n)!}
(2z)^{S-2n},
\qquad S\geq1.
\label{eq:chebyshev_monomial_definition}
\end{equation}
Comparing \eqref{eq:massless_source_sum} with
\eqref{eq:chebyshev_monomial_definition}, one finds
\begin{equation}
J\!\cdot\!\Pi^{(0)}_S\!\cdot\!J'
\;\propto\;
M^{2S}T_S(z),
\label{eq:massless_projector_chebyshev_appendix}
\end{equation}
where the proportionality constant is an overall \(S\)-dependent normalisation which can be absorbed into the cubic coupling.

Equations
\eqref{eq:massive_projector_legendre_appendix} and
\eqref{eq:massless_projector_chebyshev_appendix} reproduce the same residue diagnostic found in the on-shell analysis:
\begin{equation}
\text{strictly massless exchange: } T_S(z),
\qquad
m\to0 \text{ massive exchange: } P_S(z).
\label{eq:source_source_cheb_leg_summary}
\end{equation}
The covariant current computation, therefore, provides a direct projector-level cross-check of the Chebyshev--Legendre discontinuity.

\subsection{Spin two and the vDVZ trace mismatch}

For \(S=2\), the preceding discussion reduces to the usual vDVZ trace mismatch.
The strictly massless spin-two current exchange is
\begin{equation}
J\!\cdot\!\Pi^{(0)}_2\!\cdot\!J'
=
J_{\mu\nu}J'^{\mu\nu}
-\frac12 JJ',
\label{eq:spin2_massless_exchange_appendix}
\end{equation}
while the \(m\to0\) limit of the massive spin-two exchange is
\begin{equation}
J\!\cdot\!\Pi^{(m)}_2\!\cdot\!J'
\xrightarrow{m\to0}
J_{\mu\nu}J'^{\mu\nu}
-\frac13 JJ',
\label{eq:spin2_massive_exchange_appendix}
\end{equation}
where \(J=J^\mu{}_\mu\). For the probe current
\begin{equation}
J_{\mu\nu}=K_\mu K_\nu,
\qquad
J'_{\mu\nu}=L_\mu L_\nu,
\end{equation}
one has
\begin{equation}
J_{\mu\nu}J'^{\mu\nu}
=(K\!\cdot L)^2
=M^4z^2,
\qquad
JJ'=K^2L^2=M^4.
\end{equation}
Thus
\begin{align}
J\!\cdot\!\Pi^{(0)}_2\!\cdot\!J'
&=
M^4\left(z^2-\frac12\right)
=
\frac{M^4}{2}T_2(z),
\label{eq:spin2_chebyshev_check}
\\
J\!\cdot\!\Pi^{(m)}_2\!\cdot\!J'
&=
M^4\left(z^2-\frac13\right)
\propto
M^4P_2(z),
\label{eq:spin2_legendre_check}
\end{align}
up to the conventional normalisation of the massive coupling. 
The familiar
difference between the trace weights \(-1/2\) and \(-1/3\) is the spin-two member of the general Chebyshev--Legendre pattern.

\subsection{Relativistic versus non-relativistic sources}
\label{app:rel_vs_non-rel}

The trace-sector viewpoint also makes the physical content of the discontinuity more transparent. 
For a spin-\(S\) exchange, the source is a symmetric rank-\(S\) current \(J_{\mu_1\cdots\mu_S}\). The strictly massless and \(m\to0\) massive projectors act on the same trace decomposition, but with different weights:
\begin{align}
J\!\cdot\!\Pi_S^{(0,m\to 0)}\!\cdot\!J'
&=
\sum_{n=0}^{\lfloor S/2\rfloor}
c^{(0,m\to 0)}_{S,n}\,
J^{[n]}\!\cdot J'^{[n]}.
\end{align}
The discontinuity is precisely that the trace-sector weights differ $c^{(0)}_{S,n}\neq c^{(m\to 0)}_{S,n}$. Equivalently, the strictly massless exchange keeps only the transverse helicity-\(S\) sector, while the \(m\to0\) massive exchange retains additional lower-helicity sectors.

This difference can be probed by comparing sources with different relativistic character. 
A non-relativistic source is dominated by its time-like components.
Schematically, for a rank-\(S\) current built from the matter momentum,
\[
J_{0\cdots0}\sim M^S\rho,
\qquad
J_{i0\cdots0}\sim v\,M^S\rho,
\qquad
J_{ij0\cdots0}\sim v^2M^S\rho,
\qquad \ldots .
\]
The traces of the current are obtained by contracting pairs of indices with the metric. The first trace contains, schematically,
\[
J^{[1]}_{\mu_1\cdots\mu_{S-2}}
=
\eta^{\alpha\beta}
J_{\alpha\beta\mu_1\cdots\mu_{S-2}}
=
J_{00\mu_1\cdots\mu_{S-2}}
-
\delta^{ij}J_{ij\mu_1\cdots\mu_{S-2}},
\]
up to the sign convention of the metric. For a non relativistic source, the time-like term dominates, so
\[
J^{[1]}_{0\cdots0}
\sim
M^S\rho
\]
rather than being velocity-suppressed. 
Similarly, higher traces are also generically non-zero. 
Thus, non-relativistic matter is trace-rich: it couples not only to the transverse spin-\(S\) sector, but also efficiently to the trace sectors of the projector. Since these trace sectors are weighted differently by
\(\Pi_S^{(0)}\) and \(\Pi_S^{(m)}\), non-relativistic sources are sensitive to the extra lower-helicity content retained in the \(m\to0\) massive exchange. 

By contrast, a highly relativistic or approximately traceless source has suppressed trace components,
\[
J^{[n]}\simeq0,
\qquad n\geq1,
\]
or at least couples dominantly to the transverse/traceless part of the exchange.
Such a source probes a different linear combination of the projector trace sectors. 
Therefore, if one fixes the coupling by matching the force between non-relativistic sources, the prediction for relativistic or trace-poor sources need not agree between the strictly massless theory and the \(m\to0\) massive theory.

The familiar spin-two vDVZ discontinuity is the simplest realisation of this logic. For spin two, the source is the stress tensor \(T_{\mu\nu}\), and the two
exchange amplitudes are
\begin{align}
\mathcal A_0
&\propto
T_{\mu\nu}T'^{\mu\nu}
-\frac12 TT',
\\
\mathcal A_m
&\xrightarrow{m\to0}
T_{\mu\nu}T'^{\mu\nu}
-\frac13 TT' .
\end{align}
For non-relativistic matter,
\[
T_{00}\simeq \rho,
\qquad
T_{ij}\ll T_{00},
\qquad
T\simeq \rho,
\]
so the trace contribution is of the same order as the leading density-density term.
Hence the difference between the coefficients \(-1/2\) and \(-1/3\) affects the Newtonian force. 
One may choose the massive coupling so that this
Newtonian force agrees with the massless result. However, for radiation or any approximately traceless relativistic source,
\[
T\simeq0.
\]
The trace term then drops out, and the same coupling rescaling no longer makes the two predictions agree. This is the standard vDVZ statement: matching the
force between non-relativistic sources leaves a mismatch for relativistic probes such as light.

For higher spins, the same mechanism is encoded in the full tower of trace sectors. Matching one class of trace-rich sources fixes one particular weighted
sum of the coefficients \(c_{S,n}\). 
A trace-poor or more relativistic source projects onto a different weighted sum. 
Since the massive and strictly massless projectors have different trace-sector weights, the two theories cannot, in general, be made equivalent for all source types by a single coupling normalisation. 
The Chebyshev--Legendre residue mismatch is the compact polynomial form of this statement.

\end{appendices}

\newpage


\bibliographystyle{JHEP}
\bibliography{Refs-vJ}

@article{McGady:2013sga,
    author = "McGady, David A. and Rodina, Laurentiu",
    title = "{Higher-spin massless $S$-matrices in four-dimensions}",
    eprint = "1311.2938",
    archivePrefix = "arXiv",
    primaryClass = "hep-th",
    reportNumber = "PUPT-2454",
    doi = "10.1103/PhysRevD.90.084048",
    journal = "Phys. Rev. D",
    volume = "90",
    number = "8",
    pages = "084048",
    year = "2014"
}

@article{Benincasa:2007xk,
    author = "Benincasa, Paolo and Cachazo, Freddy",
    title = "{Consistency Conditions on the S-Matrix of Massless Particles}",
    eprint = "0705.4305",
    archivePrefix = "arXiv",
    primaryClass = "hep-th",
    reportNumber = "UWO-TH-07-09",
    month = "5",
    year = "2007"
}

@article{Schuster:2008nh,
    author = "Schuster, Philip C. and Toro, Natalia",
    title = "{Constructing the Tree-Level Yang-Mills S-Matrix Using Complex Factorization}",
    eprint = "0811.3207",
    archivePrefix = "arXiv",
    primaryClass = "hep-th",
    reportNumber = "SLAC-PUB-13426, SU-ITP-08-32",
    doi = "10.1088/1126-6708/2009/06/079",
    journal = "JHEP",
    volume = "06",
    pages = "079",
    year = "2009"
}

@article{Weinberg:1964ew,
    author = "Weinberg, Steven",
    title = "{Photons and Gravitons in  $S$-Matrix Theory: Derivation of Charge Conservation and Equality of Gravitational and Inertial Mass}",
    doi = "10.1103/PhysRev.135.B1049",
    journal = "Phys. Rev.",
    volume = "135",
    pages = "B1049--B1056",
    year = "1964"
}

@article{Aragone:1979hx,
    author = "Aragone, C. and Deser, Stanley",
    title = "{Consistency Problems of Hypergravity}",
    reportNumber = "Print-79-0377 (BRANDEIS)",
    doi = "10.1016/0370-2693(79)90808-6",
    journal = "Phys. Lett. B",
    volume = "86",
    pages = "161--163",
    year = "1979"
}

@article{Coleman:1967ad,
    author = "Coleman, Sidney R. and Mandula, J.",
    editor = "Zichichi, A.",
    title = "{All Possible Symmetries of the S Matrix}",
    doi = "10.1103/PhysRev.159.1251",
    journal = "Phys. Rev.",
    volume = "159",
    pages = "1251--1256",
    year = "1967"
}

@article{Porrati:2008rm,
    author = "Porrati, M.",
    title = "{Universal Limits on Massless High-Spin Particles}",
    eprint = "0804.4672",
    archivePrefix = "arXiv",
    primaryClass = "hep-th",
    doi = "10.1103/PhysRevD.78.065016",
    journal = "Phys. Rev. D",
    volume = "78",
    pages = "065016",
    year = "2008"
}

@inproceedings{Porrati:2012rd,
    author = "Porrati, M.",
    title = "{Old and New No Go Theorems on Interacting Massless Particles in Flat Space}",
    booktitle = "{17th International Seminar on High Energy Physics}",
    eprint = "1209.4876",
    archivePrefix = "arXiv",
    primaryClass = "hep-th",
    month = "9",
    year = "2012"
}

@article{Weinberg:1980kq,
    author = "Weinberg, Steven and Witten, Edward",
    title = "{Limits on Massless Particles}",
    reportNumber = "HUTP-80/A056",
    doi = "10.1016/0370-2693(80)90212-9",
    journal = "Phys. Lett. B",
    volume = "96",
    pages = "59--62",
    year = "1980"
}

@article{Henneaux:2012wg,
    author = "Henneaux, Marc and Lucena G{\'o}mez, Gustavo and Rahman, Rakibur",
    title = "{Higher-Spin Fermionic Gauge Fields and Their Electromagnetic Coupling}",
    eprint = "1206.1048",
    archivePrefix = "arXiv",
    primaryClass = "hep-th",
    reportNumber = "ESI-PREPRINT-NUMBER:-2380",
    doi = "10.1007/JHEP08(2012)093",
    journal = "JHEP",
    volume = "08",
    pages = "093",
    year = "2012"
}

@article{Henneaux:2013gba,
    author = "Henneaux, Marc and Lucena G{\'o}mez, Gustavo and Rahman, Rakibur",
    title = "{Gravitational Interactions of Higher-Spin Fermions}",
    eprint = "1310.5152",
    archivePrefix = "arXiv",
    primaryClass = "hep-th",
    doi = "10.1007/JHEP01(2014)087",
    journal = "JHEP",
    volume = "01",
    pages = "087",
    year = "2014"
}

@article{Hinterbichler:2017qcl,
    author = "Hinterbichler, Kurt and Joyce, Austin and Rosen, Rachel A.",
    title = "{Eikonal scattering and asymptotic superluminality of massless higher spin fields}",
    eprint = "1712.10021",
    archivePrefix = "arXiv",
    primaryClass = "hep-th",
    doi = "10.1103/PhysRevD.97.125019",
    journal = "Phys. Rev. D",
    volume = "97",
    number = "12",
    pages = "125019",
    year = "2018"
}

@article{Berends:1984rq,
    author = "Berends, Frits A. and Burgers, G. J. H. and van Dam, H.",
    title = "{On the Theoretical Problems in Constructing Interactions Involving Higher Spin Massless Particles}",
    reportNumber = "IFP-234-UNC",
    doi = "10.1016/0550-3213(85)90074-4",
    journal = "Nucl. Phys. B",
    volume = "260",
    pages = "295--322",
    year = "1985"
}

@article{Porrati:2008ha,
    author = "Porrati, Massimo and Rahman, Rakibur",
    title = "{A Model Independent Ultraviolet Cutoff for Theories with Charged Massive Higher Spin Fields}",
    eprint = "0812.4254",
    archivePrefix = "arXiv",
    primaryClass = "hep-th",
    doi = "10.1016/j.nuclphysb.2009.02.010",
    journal = "Nucl. Phys. B",
    volume = "814",
    pages = "370--404",
    year = "2009"
}

@article{Bellazzini:2017fep,
    author = "Bellazzini, Brando and Riva, Francesco and Serra, Javi and Sgarlata, Francesco",
    title = "{Beyond Positivity Bounds and the Fate of Massive Gravity}",
    eprint = "1710.02539",
    archivePrefix = "arXiv",
    primaryClass = "hep-th",
    reportNumber = "CERN-TH-2017-201",
    doi = "10.1103/PhysRevLett.120.161101",
    journal = "Phys. Rev. Lett.",
    volume = "120",
    number = "16",
    pages = "161101",
    year = "2018"
}

@article{Bonifacio:2018aon,
    author = "Bonifacio, James and Hinterbichler, Kurt",
    title = "{Universal bound on the strong coupling scale of a gravitationally coupled massive spin-2 particle}",
    eprint = "1806.10607",
    archivePrefix = "arXiv",
    primaryClass = "hep-th",
    doi = "10.1103/PhysRevD.98.085006",
    journal = "Phys. Rev. D",
    volume = "98",
    number = "8",
    pages = "085006",
    year = "2018"
}

@article{Afkhami-Jeddi:2018apj,
    author = "Afkhami-Jeddi, Nima and Kundu, Sandipan and Tajdini, Amirhossein",
    title = "{A Bound on Massive Higher Spin Particles}",
    eprint = "1811.01952",
    archivePrefix = "arXiv",
    primaryClass = "hep-th",
    doi = "10.1007/JHEP04(2019)056",
    journal = "JHEP",
    volume = "04",
    pages = "056",
    year = "2019"
}

@article{Bellazzini:2019bzh,
    author = "Bellazzini, Brando and Riva, Francesco and Serra, Javi and Sgarlata, Francesco",
    title = "{Massive Higher Spins: Effective Theory and Consistency}",
    eprint = "1903.08664",
    archivePrefix = "arXiv",
    primaryClass = "hep-th",
    doi = "10.1007/JHEP10(2019)189",
    journal = "JHEP",
    volume = "10",
    pages = "189",
    year = "2019"
}

@article{Bellazzini:2023nqj,
    author = "Bellazzini, Brando and Isabella, Giulia and Ricossa, Sergio and Riva, Francesco",
    title = "{Massive gravity is not positive}",
    eprint = "2304.02550",
    archivePrefix = "arXiv",
    primaryClass = "hep-th",
    doi = "10.1103/PhysRevD.109.024051",
    journal = "Phys. Rev. D",
    volume = "109",
    number = "2",
    pages = "024051",
    year = "2024"
}

@article{Caron-Huot:2016icg,
    author = "Caron-Huot, Simon and Komargodski, Zohar and Sever, Amit and Zhiboedov, Alexander",
    title = "{Strings from Massive Higher Spins: The Asymptotic Uniqueness of the Veneziano Amplitude}",
    eprint = "1607.04253",
    archivePrefix = "arXiv",
    primaryClass = "hep-th",
    doi = "10.1007/JHEP10(2017)026",
    journal = "JHEP",
    volume = "10",
    pages = "026",
    year = "2017"
}

@article{Bonifacio:2018vzv,
    author = "Bonifacio, James and Hinterbichler, Kurt",
    title = "{Bounds on Amplitudes in Effective Theories with Massive Spinning Particles}",
    eprint = "1804.08686",
    archivePrefix = "arXiv",
    primaryClass = "hep-th",
    doi = "10.1103/PhysRevD.98.045003",
    journal = "Phys. Rev. D",
    volume = "98",
    number = "4",
    pages = "045003",
    year = "2018"
}

@article{Arkani-Hamed:2002bjr,
    author = "Arkani-Hamed, Nima and Georgi, Howard and Schwartz, Matthew D.",
    title = "{Effective field theory for massive gravitons and gravity in theory space}",
    eprint = "hep-th/0210184",
    archivePrefix = "arXiv",
    reportNumber = "HUTP-02-A051",
    doi = "10.1016/S0003-4916(03)00068-X",
    journal = "Annals Phys.",
    volume = "305",
    pages = "96--118",
    year = "2003"
}

@article{Porrati:2008gv,
    author = "Porrati, Massimo and Rahman, Rakibur",
    title = "{Intrinsic Cutoff and Acausality for Massive Spin 2 Fields Coupled to Electromagnetism}",
    eprint = "0801.2581",
    archivePrefix = "arXiv",
    primaryClass = "hep-th",
    doi = "10.1016/j.nuclphysb.2008.05.013",
    journal = "Nucl. Phys. B",
    volume = "801",
    pages = "174--186",
    year = "2008"
}

@article{Cucchieri:1994tx,
    author = "Cucchieri, A. and Porrati, M. and Deser, Stanley",
    title = "{Tree level unitarity constraints on the gravitational couplings of higher spin massive fields}",
    eprint = "hep-th/9408073",
    archivePrefix = "arXiv",
    reportNumber = "IASSNS-HEP-94-63, NYU-TH-94-05-01A, CERN-TH-7388-94",
    doi = "10.1103/PhysRevD.51.4543",
    journal = "Phys. Rev. D",
    volume = "51",
    pages = "4543--4549",
    year = "1995"
}

@article{Fradkin:1987ks,
    author = "Fradkin, E. S. and Vasiliev, Mikhail A.",
    title = "{On the Gravitational Interaction of Massless Higher Spin Fields}",
    doi = "10.1016/0370-2693(87)91275-5",
    journal = "Phys. Lett. B",
    volume = "189",
    pages = "89--95",
    year = "1987"
}

@article{Vasiliev:1990en,
    author = "Vasiliev, Mikhail A.",
    title = "{Consistent equation for interacting gauge fields of all spins in (3+1)-dimensions}",
    reportNumber = "LEBEDEV-90-29",
    doi = "10.1016/0370-2693(90)91400-6",
    journal = "Phys. Lett. B",
    volume = "243",
    pages = "378--382",
    year = "1990"
}

@article{Vasiliev:1999ba,
    author = "Vasiliev, Mikhail A.",
    editor = "Shifman, Mikhail A.",
    title = "{Higher spin gauge theories: Star product and AdS space}",
    eprint = "hep-th/9910096",
    archivePrefix = "arXiv",
    reportNumber = "FIAN-TD-24-99",
    doi = "10.1142/9789812793850_0030",
    pages = "533--610",
    month = "10",
    year = "1999"
}

@article{Vasiliev:2003cph,
    author = "Vasiliev, M. A.",
    editor = "Ivanov, Evgeny and Pashnev, Anatoly",
    title = "{Higher spin gauge theories in various dimensions}",
    eprint = "hep-th/0401177",
    archivePrefix = "arXiv",
    doi = "10.1002/prop.200410167",
    journal = "PoS",
    volume = "JHW2003",
    pages = "003",
    year = "2003"
}

@article{Porrati:2000cp,
    author = "Porrati, M.",
    title = "{No van Dam-Veltman-Zakharov discontinuity in AdS space}",
    eprint = "hep-th/0011152",
    archivePrefix = "arXiv",
    reportNumber = "NYU-TH-00-11-02",
    doi = "10.1016/S0370-2693(00)01380-0",
    journal = "Phys. Lett. B",
    volume = "498",
    pages = "92--96",
    year = "2001"
}

@article{Kogan:2000uy,
    author = "Kogan, Ian I. and Mouslopoulos, Stavros and Papazoglou, Antonios",
    title = "{The m ---{\ensuremath{>}} 0 limit for massive graviton in dS(4) and AdS(4): How to circumvent the van Dam-Veltman-Zakharov discontinuity}",
    eprint = "hep-th/0011138",
    archivePrefix = "arXiv",
    reportNumber = "OUTP-00-47P",
    doi = "10.1016/S0370-2693(01)00209-X",
    journal = "Phys. Lett. B",
    volume = "503",
    pages = "173--180",
    year = "2001"
}

@article{Anninos:2011ui,
    author = "Anninos, Dionysios and Hartman, Thomas and Strominger, Andrew",
    title = "{Higher Spin Realization of the dS/CFT Correspondence}",
    eprint = "1108.5735",
    archivePrefix = "arXiv",
    primaryClass = "hep-th",
    doi = "10.1088/1361-6382/34/1/015009",
    journal = "Class. Quant. Grav.",
    volume = "34",
    number = "1",
    pages = "015009",
    year = "2017"
}

@article{Karch:2001jb,
    author = "Karch, Andreas and Katz, Emanuel and Randall, Lisa",
    title = "{Absence of a VVDZ discontinuity in AdS(AdS)}",
    eprint = "hep-th/0106261",
    archivePrefix = "arXiv",
    reportNumber = "MIT-CTP-3148",
    doi = "10.1088/1126-6708/2001/12/016",
    journal = "JHEP",
    volume = "12",
    pages = "016",
    year = "2001"
}

@article{Arkani-Hamed:2015bza,
    author = "Arkani-Hamed, Nima and Maldacena, Juan",
    title = "{Cosmological Collider Physics}",
    eprint = "1503.08043",
    archivePrefix = "arXiv",
    primaryClass = "hep-th",
    month = "3",
    year = "2015"
}

@article{Anninos:2019nib,
    author = "Anninos, D. and De Luca, V. and Franciolini, G. and Kehagias, A. and Riotto, A.",
    title = "{Cosmological Shapes of Higher-Spin Gravity}",
    eprint = "1902.01251",
    archivePrefix = "arXiv",
    primaryClass = "hep-th",
    doi = "10.1088/1475-7516/2019/04/045",
    journal = "JCAP",
    volume = "04",
    pages = "045",
    year = "2019"
}

@article{vanDam:1970vg,
    author = "van Dam, H. and Veltman, M. J. G.",
    title = "{Massive and massless Yang-Mills and gravitational fields}",
    doi = "10.1016/0550-3213(70)90416-5",
    journal = "Nucl. Phys. B",
    volume = "22",
    pages = "397--411",
    year = "1970"
}

@article{Zakharov:1970cc,
    author = "Zakharov, V. I.",
    title = "{Linearized gravitation theory and the graviton mass}",
    journal = "JETP Lett.",
    volume = "12",
    pages = "312",
    year = "1970"
}

@article{Boulware:1972yco,
    author = "Boulware, D. G. and Deser, Stanley",
    title = "{Can gravitation have a finite range?}",
    doi = "10.1103/PhysRevD.6.3368",
    journal = "Phys. Rev. D",
    volume = "6",
    pages = "3368--3382",
    year = "1972"
}

@article{Hinterbichler:2011tt,
    author = "Hinterbichler, Kurt",
    title = "{Theoretical Aspects of Massive Gravity}",
    eprint = "1105.3735",
    archivePrefix = "arXiv",
    primaryClass = "hep-th",
    doi = "10.1103/RevModPhys.84.671",
    journal = "Rev. Mod. Phys.",
    volume = "84",
    pages = "671--710",
    year = "2012"
}

@article{Berends:1979je, 
    author = "Berends, Frits A. and van Reisen, J. C. J. M.",
    title = "{On Spin 3 Field Theory and the Zero Mass Limit of Higher Spin Theories}",
    reportNumber = "Print-79-0604 (LEIDEN)",
    doi = "10.1016/0550-3213(80)90512-X",
    journal = "Nucl. Phys. B",
    volume = "164",
    pages = "286--304",
    year = "1980"
}

@article{Francia:2007qt,
    author = "Francia, Dario and Mourad, J. and Sagnotti, A.",
    title = "{Current Exchanges and Unconstrained Higher Spins}",
    eprint = "hep-th/0701163",
    archivePrefix = "arXiv",
    doi = "10.1016/j.nuclphysb.2007.03.021",
    journal = "Nucl. Phys. B",
    volume = "773",
    pages = "203--237",
    year = "2007"
}

@article{Francia:2008hd,
    author = "Francia, Dario and Mourad, J. and Sagnotti, A.",
    title = "{(A)dS exchanges and partially-massless higher spins}",
    eprint = "0803.3832",
    archivePrefix = "arXiv",
    primaryClass = "hep-th",
    doi = "10.1016/j.nuclphysb.2008.04.023",
    journal = "Nucl. Phys. B",
    volume = "804",
    pages = "383--420",
    year = "2008"
}

@article{Sagnotti:2010jt,
    author = "Sagnotti, A.",
    editor = "Anagnostopoulos, K. and Antoniadis, I. and Bahns, D. and Irges, N. and Kehagias, A. and Lazarides, G. and Luest, Dieter and Steinacker, Harold and Zoupanos, George",
    title = "{Higher Spins and Current Exchanges}",
    eprint = "1002.3388",
    archivePrefix = "arXiv",
    primaryClass = "hep-th",
    doi = "10.22323/1.155.0106",
    journal = "PoS",
    volume = "CORFU2011",
    pages = "106",
    year = "2011"
}

@article{Vainshtein:1972sx,
    author = "Vainshtein, A. I.",
    title = "{To the problem of nonvanishing gravitation mass}",
    doi = "10.1016/0370-2693(72)90147-5",
    journal = "Phys. Lett. B",
    volume = "39",
    pages = "393--394",
    year = "1972"
}

@article{Porrati:2002cp,
    author = "Porrati, M.",
    title = "{Fully covariant van Dam-Veltman-Zakharov discontinuity, and absence thereof}",
    eprint = "hep-th/0203014",
    archivePrefix = "arXiv",
    reportNumber = "NYU-TH-02-02-01",
    doi = "10.1016/S0370-2693(02)01656-8",
    journal = "Phys. Lett. B",
    volume = "534",
    pages = "209--215",
    year = "2002"
}

@article{Deffayet:2001uk,
    author = "Deffayet, Cedric and Dvali, G. R. and Gabadadze, Gregory and Vainshtein, Arkady I.",
    title = "{Nonperturbative continuity in graviton mass versus perturbative discontinuity}",
    eprint = "hep-th/0106001",
    archivePrefix = "arXiv",
    reportNumber = "NYU-TH-01-05-15, TPI-MINN-01-23, UMN-TH-2009-01",
    doi = "10.1103/PhysRevD.65.044026",
    journal = "Phys. Rev. D",
    volume = "65",
    pages = "044026",
    year = "2002"
}

@article{Bekaert:2010hw,
    author = "Bekaert, Xavier and Boulanger, Nicolas and Sundell, Per",
    title = "{How higher-spin gravity surpasses the spin two barrier: no-go theorems versus yes-go examples}",
    eprint = "1007.0435",
    archivePrefix = "arXiv",
    primaryClass = "hep-th",
    doi = "10.1103/RevModPhys.84.987",
    journal = "Rev. Mod. Phys.",
    volume = "84",
    pages = "987--1009",
    year = "2012"
}

@article{Rahman:2012thy,
    author = "Rahman, Rakibur",
    title = "{Higher Spin Theory - Part I}",
    eprint = "1307.3199",
    archivePrefix = "arXiv",
    primaryClass = "hep-th",
    doi = "10.22323/1.195.0004",
    journal = "PoS",
    volume = "ModaveVIII",
    pages = "004",
    year = "2012"
}

@article{Rahman:2015pzl,
    author = "Rahman, Rakibur and Taronna, Massimo",
    title = "{From Higher Spins to Strings: A Primer}",
    eprint = "1512.07932",
    archivePrefix = "arXiv",
    primaryClass = "hep-th",
    doi = "10.1007/978-3-031-59656-8_1",
    journal = "Lect. Notes Phys.",
    volume = "1028",
    pages = "1--119",
    year = "2024"
}

@article{Arkani-Hamed:2017jhn,
    author = "Arkani-Hamed, Nima and Huang, Tzu-Chen and Huang, Yu-tin",
    title = "{Scattering amplitudes for all masses and spins}",
    eprint = "1709.04891",
    archivePrefix = "arXiv",
    primaryClass = "hep-th",
    reportNumber = "NCTS-TH/1714, NCTS-TH-1714",
    doi = "10.1007/JHEP11(2021)070",
    journal = "JHEP",
    volume = "11",
    pages = "070",
    year = "2021"
}

@article{Durieux:2019eor,
    author = "Durieux, Gauthier and Kitahara, Teppei and Shadmi, Yael and Weiss, Yaniv",
    title = "{The electroweak effective field theory from on-shell amplitudes}",
    eprint = "1909.10551",
    archivePrefix = "arXiv",
    primaryClass = "hep-ph",
    doi = "10.1007/JHEP01(2020)119",
    journal = "JHEP",
    volume = "01",
    pages = "119",
    year = "2020"
}

@article{Dreiner:2008tw,
    author = "Dreiner, Herbi K. and Haber, Howard E. and Martin, Stephen P.",
    title = "{Two-component spinor techniques and Feynman rules for quantum field theory and supersymmetry}",
    eprint = "0812.1594",
    archivePrefix = "arXiv",
    primaryClass = "hep-ph",
    reportNumber = "BN-TH-2008-12, SCIPP-08-08, FERMILAB-PUB-09-855-T, BN-TH-2008-12 and SCIPP-08/08",
    doi = "10.1016/j.physrep.2010.05.002",
    journal = "Phys. Rept.",
    volume = "494",
    pages = "1--196",
    year = "2010"
}

@article{Witten:2003nn,
    author = "Witten, Edward",
    title = "{Perturbative gauge theory as a string theory in twistor space}",
    eprint = "hep-th/0312171",
    archivePrefix = "arXiv",
    doi = "10.1007/s00220-004-1187-3",
    journal = "Commun. Math. Phys.",
    volume = "252",
    pages = "189--258",
    year = "2004"
}

@article{Elvang:2013cua, 
    author = "Elvang, Henriette and Huang, Yu-tin",
    title = "{Scattering Amplitudes}",
    eprint = "1308.1697",
    archivePrefix = "arXiv",
    primaryClass = "hep-th",
    month = "8",
    year = "2013"
}

@article{Chung:2018kqs,
    author = "Chung, Ming-Zhi and Huang, Yu-Tin and Kim, Jung-Wook and Lee, Sangmin",
    title = "{The simplest massive S-matrix: from minimal coupling to Black Holes}",
    eprint = "1812.08752",
    archivePrefix = "arXiv",
    primaryClass = "hep-th",
    reportNumber = "NCTS-TH/1817",
    doi = "10.1007/JHEP04(2019)156",
    journal = "JHEP",
    volume = "04",
    pages = "156",
    year = "2019"
}

@article{Chandrasekaran_2018,
   title={Higher-point positivity},
   volume={2018},
   ISSN={1029-8479},
   url={http://dx.doi.org/10.1007/JHEP11(2018)015},
   DOI={10.1007/jhep11(2018)015},
   number={11},
   journal={Journal of High Energy Physics},
   publisher={Springer Science and Business Media LLC},
   author={Chandrasekaran, Venkatesa and Remmen, Grant N. and Shahbazi-Moghaddam, Arvin},
   year={2018},
   month=nov }

@article{Falkowski:2020mjq,
    author = "Falkowski, Adam and Isabella, Giulia",
    title = "{Matter coupling in massive gravity}",
    eprint = "2001.06800",
    archivePrefix = "arXiv",
    primaryClass = "hep-th",
    doi = "10.1007/JHEP04(2020)014",
    journal = "JHEP",
    volume = "04",
    pages = "014",
    year = "2020"
}

@article{Falkowski:2020fsu,
    author = "Falkowski, Adam and Isabella, Giulia and Machado, Camila S.",
    title = "{On-shell effective theory for higher-spin dark matter}",
    eprint = "2011.05339",
    archivePrefix = "arXiv",
    primaryClass = "hep-ph",
    reportNumber = "DESY-20-198",
    doi = "10.21468/SciPostPhys.10.5.101",
    journal = "SciPost Phys.",
    volume = "10",
    number = "5",
    pages = "101",
    year = "2021"
}

@article{Balkin:2021dko,
    author = "Balkin, Reuven and Durieux, Gauthier and Kitahara, Teppei and Shadmi, Yael and Weiss, Yaniv",
    title = "{On-shell Higgsing for EFTs}",
    eprint = "2112.09688",
    archivePrefix = "arXiv",
    primaryClass = "hep-ph",
    reportNumber = "CERN-TH-2021-212",
    doi = "10.1007/JHEP03(2022)129",
    journal = "JHEP",
    volume = "03",
    pages = "129",
    year = "2022"
}

@article{Arkani-Hamed:2019ymq,
    author = "Arkani-Hamed, Nima and Huang, Yu-tin and O'Connell, Donal",
    title = "{Kerr black holes as elementary particles}",
    eprint = "1906.10100",
    archivePrefix = "arXiv",
    primaryClass = "hep-th",
    reportNumber = "NCTS-TH/1905",
    doi = "10.1007/JHEP01(2020)046",
    journal = "JHEP",
    volume = "01",
    pages = "046",
    year = "2020"
}

@article{Guevara:2019fsj,
    author = "Guevara, Alfredo and Ochirov, Alexander and Vines, Justin",
    title = "{Black-hole scattering with general spin directions from minimal-coupling amplitudes}",
    eprint = "1906.10071",
    archivePrefix = "arXiv",
    primaryClass = "hep-th",
    doi = "10.1103/PhysRevD.100.104024",
    journal = "Phys. Rev. D",
    volume = "100",
    number = "10",
    pages = "104024",
    year = "2019"
}

@article{Aoude:2020ygw,
    author = "Aoude, Rafael and Haddad, Kays and Helset, Andreas",
    title = "{Tidal effects for spinning particles}",
    eprint = "2012.05256",
    archivePrefix = "arXiv",
    primaryClass = "hep-th",
    doi = "10.1007/JHEP03(2021)097",
    journal = "JHEP",
    volume = "03",
    pages = "097",
    year = "2021"
}

@article{Csaki:2020inw,
    author = "Csaki, Csaba and Hong, Sungwoo and Shirman, Yuri and Telem, Ofri and Terning, John and Waterbury, Michael",
    title = "{Scattering amplitudes for monopoles: pairwise little group and pairwise helicity}",
    eprint = "2009.14213",
    archivePrefix = "arXiv",
    primaryClass = "hep-th",
    doi = "10.1007/JHEP08(2021)029",
    journal = "JHEP",
    volume = "08",
    pages = "029",
    year = "2021"
}

@article{Cembranos:2017vgi,
    author = "Cembranos, J. A. R. and Maroto, A. L. and Villarrubia-Rojo, H.",
    title = "{Constraints on hidden gravitons from fifth-force experiments and stellar energy loss}",
    eprint = "1706.07818",
    archivePrefix = "arXiv",
    primaryClass = "hep-ph",
    doi = "10.1007/JHEP09(2017)104",
    journal = "JHEP",
    volume = "09",
    pages = "104",
    year = "2017"
}

@article{Camanho:2014apa,
    author = "Camanho, Xian O. and Edelstein, Jose D. and Maldacena, Juan and Zhiboedov, Alexander",
    title = "{Causality Constraints on Corrections to the Graviton Three-Point Coupling}",
    eprint = "1407.5597",
    archivePrefix = "arXiv",
    primaryClass = "hep-th",
    doi = "10.1007/JHEP02(2016)020",
    journal = "JHEP",
    volume = "02",
    pages = "020",
    year = "2016"
}

@article{Ruhdorfer:2019qmk,
    author = "Ruhdorfer, Maximilian and Serra, Javi and Weiler, Andreas",
    title = "{Effective Field Theory of Gravity to All Orders}",
    eprint = "1908.08050",
    archivePrefix = "arXiv",
    primaryClass = "hep-ph",
    reportNumber = "TUM-HEP-1205-19",
    doi = "10.1007/JHEP05(2020)083",
    journal = "JHEP",
    volume = "05",
    pages = "083",
    year = "2020"
}

@article{Caron-Huot:2022ugt,
    author = "Caron-Huot, Simon and Li, Yue-Zhou and Parra-Martinez, Julio and Simmons-Duffin, David",
    title = "{Causality constraints on corrections to Einstein gravity}",
    eprint = "2201.06602",
    archivePrefix = "arXiv",
    primaryClass = "hep-th",
    reportNumber = "CALT-TH 2021-003",
    doi = "10.1007/JHEP05(2023)122",
    journal = "JHEP",
    volume = "05",
    pages = "122",
    year = "2023"
}

@article{Cheung:2016yqr,
    author = "Cheung, Clifford and Remmen, Grant N.",
    title = "{Positive Signs in Massive Gravity}",
    eprint = "1601.04068",
    archivePrefix = "arXiv",
    primaryClass = "hep-th",
    reportNumber = "CALT-TH-2015-062",
    doi = "10.1007/JHEP04(2016)002",
    journal = "JHEP",
    volume = "04",
    pages = "002",
    year = "2016"
}

@article{Aubert:2003je,
    author = "Aubert, Aurele",
    title = "{Strong coupling in massive gravity by direct calculation}",
    eprint = "hep-th/0312246",
    archivePrefix = "arXiv",
    doi = "10.1103/PhysRevD.69.087502",
    journal = "Phys. Rev. D",
    volume = "69",
    pages = "087502",
    year = "2004"
}

@article{Holstein:2008sx,
    author = "Holstein, Barry R. and Ross, Andreas",
    title = "{Spin Effects in Long Range Gravitational Scattering}",
    eprint = "0802.0716",
    archivePrefix = "arXiv",
    primaryClass = "hep-ph",
    month = "2",
    year = "2008"
}

@article{Bern:2019crd,
    author = "Bern, Zvi and Cheung, Clifford and Roiban, Radu and Shen, Chia-Hsien and Solon, Mikhail P. and Zeng, Mao",
    title = "{Black Hole Binary Dynamics from the Double Copy and Effective Theory}",
    eprint = "1908.01493",
    archivePrefix = "arXiv",
    primaryClass = "hep-th",
    reportNumber = "CERN-TH-2019-128, CALT-TH 2019-026, UCLA/TEP/2019/103",
    doi = "10.1007/JHEP10(2019)206",
    journal = "JHEP",
    volume = "10",
    pages = "206",
    year = "2019"
}

@article{Cheung:2018wkq,
    author = "Cheung, Clifford and Rothstein, Ira Z. and Solon, Mikhail P.",
    title = "{From Scattering Amplitudes to Classical Potentials in the Post-Minkowskian Expansion}",
    eprint = "1808.02489",
    archivePrefix = "arXiv",
    primaryClass = "hep-th",
    reportNumber = "CALT-TH-2018-031",
    doi = "10.1103/PhysRevLett.121.251101",
    journal = "Phys. Rev. Lett.",
    volume = "121",
    number = "25",
    pages = "251101",
    year = "2018"
}

@article{Baratella:2021guc,
    author = "Baratella, Pietro and Haslehner, Dominik and Ruhdorfer, Maximilian and Serra, Javi and Weiler, Andreas",
    title = "{RG of GR from on-shell amplitudes}",
    eprint = "2109.06191",
    archivePrefix = "arXiv",
    primaryClass = "hep-th",
    reportNumber = "TUM-HEP-1363/21",
    doi = "10.1007/JHEP03(2022)156",
    journal = "JHEP",
    volume = "03",
    pages = "156",
    year = "2022"
}

@article{Moynihan:2017tva,
    author = "Moynihan, Nathan and Murugan, Jeff",
    title = "{Comments on scattering in massive gravity, vDVZ and BCFW}",
    eprint = "1711.03956",
    archivePrefix = "arXiv",
    primaryClass = "hep-th",
    doi = "10.1088/1361-6382/aacb73",
    journal = "Class. Quant. Grav.",
    volume = "35",
    number = "15",
    pages = "155005",
    year = "2018"
}

@article{Burger:2020kle,
    author = "Burger, Daniel J. and Moynihan, Nathan and Murugan, Jeff",
    title = "{On-Shell Perspectives on the Massless Limit of Massive Supergravity}",
    eprint = "2005.14077",
    archivePrefix = "arXiv",
    primaryClass = "hep-th",
    month = "5",
    year = "2020"
}

\end{document}